\documentclass[a4paper,11pt]{article}
\usepackage[utf8]{inputenc}
\usepackage[T1]{fontenc}
\usepackage{mathtools}
\usepackage{amsfonts}
\usepackage{amsmath}
\usepackage{amssymb}
\usepackage{optidef}
\usepackage{graphicx}
\graphicspath{{./Figure}}
\usepackage{geometry}
\usepackage{url}
\usepackage{setspace}
\usepackage{multirow}
\usepackage{colortbl}
\usepackage{tabulary}
\usepackage{etoolbox}
\usepackage{booktabs}
\usepackage{subfig}
\usepackage{array}
\usepackage{caption}
\usepackage{adjustbox}
\usepackage{tabularx}
\usepackage{color}
\usepackage[table]{xcolor}
\usepackage{hyperref}
\hypersetup{
  colorlinks,
  citecolor=violet,
  linkcolor=red,
  urlcolor=blue}
\usepackage[round]{natbib}
\PassOptionsToPackage{
        natbib=true,
        style=authoryear-comp,
        hyperref=true,
        backend=biber,
        maxbibnames=99,
        firstinits=true,
        uniquename=init,
        maxcitenames=2,
        parentracker=true,
        url=false,
        doi=false,
        isbn=false,
        eprint=false,
        backref=true,
            }   {biblatex}

\newtheorem{remark}{Remark}

\DeclareMathOperator*{\argmin}{arg\,min}

\DeclarePairedDelimiter{\abs}{\lvert}{\rvert}

\begin{document}

\title{A new behavioral model for portfolio selection using the Half-Full/Half-Empty approach}

\author{F Cesarone$^1$, M Corradini$^1$, L Lampariello$^1$,  J Riccioni$^1$\thanks{Corresponding author.}\\
{\small $^1$ \emph{ Roma Tre University - Department of Business Studies}} \\
{\small \emph{Via Silvio D'Amico, 77 -- 00145, Rome, Italy}} \\
{\footnotesize francesco.cesarone@uniroma3.it, massimiliano.corradini@uniroma3.it,
}\\
{\footnotesize lorenzo.lampariello@uniroma3.it, jessica.riccioni@uniroma3.it}\\
}

\date{\today}
\maketitle

\begin{abstract}

We focus on a behavioral model, that has been recently proposed in the literature, whose rational can be traced back to the Half-Full/Half-Empty glass metaphor.
%
%
More precisely, we generalize the Half-Full/Half-Empty approach to the context of positive and negative lotteries and give financial and behavioral interpretations of the Half-Full/Half-Empty parameters.
We develop a portfolio selection model based on the Half-Full/Half-Empty strategy, resulting in a nonconvex optimization problem, which, nonetheless, is proven to be equivalent to an alternative Mixed-Integer Linear Programming formulation. 
By means of the ensuing empirical analysis, based on three real-world datasets, the Half-Full/Half-Empty model is shown to be very versatile by appropriately varying its parameters, and to provide portfolios displaying promising performances in terms of risk and profitability, compared with Prospect Theory, risk minimization approaches and Equally-Weighted portfolios.

\medskip


\end{abstract}

\section{Introduction}

Theoretical models of individual behavior in decision-making under risk conditions begun to have a solid foundation in the mid-1900s with the development of the Expected Utility Theory (EUT) by von Neumann \citep{neumann1944}.
According to EUT, if the preferences of individuals in risky decision making satisfy some fundamental axioms, the Decision Maker (DM) maximizes the expected utility function, and such choice is considered to be rational \citep{keeney1993}: in this framework, individuals are  taken to be rational agents. 
However, many criticisms have followed over the years due to the discrepancy between the theorized rational behavior and the real one, which can actually be influenced by multiple factors.

The evolution of the literature on decision-making under risk has long been driven by the discovery of paradoxes \citep{fox2014prospect,luce2008utility,wu2004decision}. Old paradoxes such as the Allais one \citep{allais1953} and its variations \citep{kahnemanandtversky1979} challenged the descriptive power of EUT. 
For this reason, over the last 60 years, decision theorists have been trying to develop a theory of choice that can give a satisfactory description of decision-makers' behavior in risky situations.

\noindent
Although several theories have been proposed over time  \citep[see, e.g.,][]{fox2014prospect,starmer2000developments,wu2004decision}, Prospect Theory (PT) \citep[][]{kahnemanandtversky1979} and Cumulative Prospect Theory (CPT) \citep[][]{tverskyandkahneman1992} have emerged as the leading descriptive paradigm for modeling decision-making under risk \citep{camerer1998bounded,fox2014prospect,starmer2000developments,wu2004decision} and have inspired a large body of theoretical and empirical work \citep{abdellaoui2000parameter,abdellaoui2002genuine,camerer1992recent,diecidue2001intuition,gonzalez1999shape,karni1987preference,luce2014utility,luce2001reduction,machina1982expected,prelec1998probability,quiggin2012generalized,schmeidler1989subjective,starmer1989violations,tversky1995risk,viscusi1989prospective,wakker1994separating,wakker1996sure,wakker2001testing,wakker1994comonotonic,wu1996curvature,wu1998common,wu1999nonlinear,yaari1987dual}. 
Despite this apparent success, several scholars have questioned the underlying assumptions and applicability of PT and CPT, which seem not to describe some empirical evidence \citep{baltussen2006violations,blavatskyy2005back,brandstatter2006priority,hertwig2004decisions,humphrey1995regret,levy2002prospect,lopes1999role,marley2005independence,neilson2002further,payne2005whether,starmer2000developments,starmer1993testing,weber1997reasons,wu1999nonlinear,wu2008empirical,wu2005testing}.

\noindent
PT and CPT are, thus, at the same time, ``one of the most confirmed and falsified decision-making models currently in circulation'' \citep{wakker2010prospect}.
However, \cite{birnbaum2008new} has recently produced systematic empirical cases that confute both PT and CPT approaches and have received considerable attention in the relevant literature \citep{fox2014prospect,luce2008utility,wakker2010prospect}.
In fact, the author has presented 11 new paradoxes showing ``self-contradiction or systematic false predictions'' \citep[][p. 463]{birnbaum2008new}, leading to the conclusion that PT and CPT cannot ``be retained as a descriptive model of decision making'' \citep[][p. 464]{birnbaum2008new}.

\noindent
The transfer of attention exchange (TAX) model ``correctly predicts results with all 11 new paradoxes'' \citep[][p. 463]{birnbaum1979source,birnbaum1997tests,birnbaum2008new}. 
But, while this model fits the data better than PT and CPT \citep{birnbaum2008new,fox2014prospect,wakker2010prospect}, it is less parsimonious \citep[][p. 408]{wu2004decision}, involving the presence of a large number of free parameters, and does not have ``the particularly tractable form of prospect theory, with psychological interpretations for its parameters'' \citep[][p. 351]{wakker2010prospect}.

In \cite{cencietal2015}, a simple decision-making under risk approach for positive lotteries has been proposed to explain the well-known ``old and new'' paradoxes at a stroke.
The rationale behind the resulting model can be easily understood through the metaphor of the ``glass half-full, glass half-empty'': different individuals may evaluate the same situation differently and are typically divided between those who focus on positive aspects (the optimists/glass half-full) and those who focus on negative ones (the pessimists/glass half-empty). 
The Half-Full/Half-Empty (HF/HE) model suggests that a DM's evaluation of a choice situation may depend on how the DM assesses the difference between outcomes above and below the mean value of the lottery considered. 
The DM's evaluation of the lottery (and the consequent attitude towards risk) depends on two parameters only: the DM's degree of optimism/pessimism, and a probability distortion parameter that determines the DM's decision weights.
In \cite{corradini2022half}, the author investigates how to suitably calibrate this model to explain ``old and new'' paradoxes, even in the case of negative lotteries (Note that, in \cite{corradini2022half}, only calibration of negative lotteries in \cite{kahnemanandtversky1979} is considered since negative lotteries in \cite{birnbaum2008new} are trivially explained).

\noindent
Compared to other decision-making models that have been recently proposed to address the paradoxes in \cite{birnbaum2008new}, the advantages of the HF/HE approach are twofold. 
On the one hand, it is very intuitive, as it is closely related to the Expected Value (EV) criterion, and the related parameters have a clear and well-established behavioral interpretation \citep[see, e.g.,][]{hurwicz1951optimality,arrow1972optimality,kahnemanandtversky1979,tversky1995risk}.
On the other hand, the HF/HE approach is highly parsimonious and flexible, providing an extremely simple way of modeling behavior, based on just three parameters.

Behavioral models for portfolio selection incorporate investor's perception of risk into the decision-making process, avoiding the pitfalls of suboptimal investment choices.
Theoretical evidence exists in the literature on risk reduction in portfolio selection models based on the behavior of DMs who are naturally loss averse \citep[]{HensandBachmann2008,grishinaetal2017}.
\noindent
\cite{benartzi1995myopic} focus their study on a particular loss aversion concept, i.e. the so-called myopic loss aversion, which is key in PT, to explain the equity premium puzzle \citep[see also][]{mehra1985equity} for DMs who evaluate their portfolios annually.
\cite{levyandlevy2003}, among the Pareto-optimal mean-variance portfolios, propose to select the ones with the highest prospect utility, in case of normally distributed asset returns. 
\cite{pirvu2012multi} generalize the results of \cite{levyandlevy2003} for elliptic symmetric distributions of risky assets. 
\cite{de2007computational} apply the PT approach to portfolio selection, assuming equally likely historical scenarios for the asset returns. 
The PT portfolios are obtained by solving a nonconvex optimization problem through a heuristic. 
Other scholars investigate portfolio selection models based on CPT
\citep[see][]{bernard2010static, he2011portfolio, consiglietal2019}.
\cite{barroetal2020} compare different portfolio selection models, including models based on PT and CPT, and approaches based on risk minimization, where risk is measured by variance and mean absolute deviation.

In this paper, we refine the approach in \cite{cencietal2015} and \cite{corradini2022half} by tailoring it to portfolio selection problems.
More precisely, we generalize the HF/HE approach to the context of mixed (i.e. positive and negative) lotteries and discuss financial and behavioral interpretation of the HF/HE-related parameters.
We develop a portfolio selection model based on the HF/HE approach that results in a nonconvex optimization problem, for which we suggest an alternative, equivalent Mixed-Integer Linear Programming (MILP) formulation. 
By means of the empirical analysis we give, based on three real-world data sets, the HF/HE model is shown to be very versatile by appropriately setting its parameters, providing portfolios that display promising performances in terms of risk and profitability, compared with Prospect Theory and risk minimization approaches and the Equally-Weighted portfolio.

The rest of the paper is organized as follows. 
In Section \ref{sec:HFHEapproach}, we describe how to apply the HF/HE approach to portfolio selection.
More in detail, in Section \ref{sec:Preliminary}, we introduce some preliminary concepts related to the decision-making model under risk in \cite{cencietal2015} and \cite{corradini2022half}.
In Section \ref{sec:PositiveNegative}, we extend the HF/HE approach to the context of positive and negative lotteries and present a detailed discussion about the financial interpretation of HF/HE parameters. 
In Section \ref{sec:HF/HEportfolioModel}, we adapt the HF/HE approach for portfolio selection purposes, obtaining a nonconvex optimization problem, for which, in Section \ref{sec:MILP}, we propose a (MILP) equivalent formulation.
In Section \ref{sec:OtherModels}, we present the PT-based portfolio selection model and two minimum-risk portfolio strategies that referred to for comparative purposes.
In Section \ref{sec:EmpiricalAnalysis}, we provide an extensive empirical analysis based on three real-world datasets, where the performance of the HF/HE portfolio is compared with that of Prospect Theory, Equally-Weighted, minimum variance and Mean Absolute Deviation (MAD) portfolios.
Conclusions are drawn in Section \ref{sec:conclusions}.

\section{Half-Full/Half-Empty approach}\label{sec:HFHEapproach}

After gaining insight, in Section \ref{sec:Preliminary}, about the decision-making model under risk that has been developed in \cite{cencietal2015} and \cite{corradini2022half}, we show in Section \ref{sec:HF/HEportfolioModel} how to tailor it to portfolio selection purposes: as for the resulting nonconvex optimization problem, we propose, in Section \ref{sec:MILP}, a MILP equivalent formulation.

\subsection{Preliminary concepts\label{sec:Preliminary}}

We recall here some preliminary notions of the HF/HE approach in \cite{cencietal2015}, where the authors consider only nonnegative random variables.

\noindent
Denoting by $Y$ a generic nonnegative lottery, 
we can express its EV as follows:
\begin{equation}
	H_{EV}[Y]=\mathbb{E}[Y]= \mu + \mathbb{E}[Y-\mu] \, ,
	\label{hfhe_ev1}
\end{equation}
where obviously $\mu=\mathbb{E}[Y]$.
Distinguishing the part of $Y$ below the mean from the part above the mean, we can write $(Y-\mu)= (Y-\mu)_+ + (Y-\mu)_{-}$, where $(Y-\mu)_+=\max\{Y-\mu,0\}$ and $(Y-\mu)_{-}=\min\{Y-\mu,0\}$. Expression \eqref{hfhe_ev1} becomes
\begin{equation}
	H_{EV}[Y]= \mu+2 \Bigl \{ \frac{1}{2} \mathbb{E}[(Y-\mu)_+] + \frac{1}{2} \mathbb{E}[(Y-\mu)_-] \Bigr \}.
	\label{hfhe_ev2}
\end{equation}
The expected value of $Y$ can be generalized by assuming that the DM assigns a weight $\lambda$ different from $\frac{1}{2}$ for realizations of $Y$ above and those below the mean, according to the DM's degree of optimism/pessimism.
Therefore, expression \eqref{hfhe_ev2} becomes
\begin{equation}
	H_\lambda [Y]= \mu+2 \bigl \{ \lambda \mathbb{E}[(Y-\mu)_+] + (1-\lambda) \mathbb{E}[(Y-\mu)_-] \bigr \}, 
	\label{hfhe_lambda}
\end{equation}
where $\lambda \in [0,1]$ and represents the DM's degree of optimism/pessimism.
More precisely, an \emph{optimistic} DM sets $\lambda \in (\frac{1}{2},1]$, thus assigning a weight, that is larger than $\frac{1}{2}$, to the average of the values of $Y$ above $\mu$. 
If a DM is \emph{pessimistic}, then $\lambda \in [0,\frac{1}{2})$, i.e. the DM attributes a weight, that is larger than $\frac{1}{2}$, to the average of the values of $Y$ below $\mu$.
If $\lambda=\frac{1}{2}$, \eqref{hfhe_lambda} coincides with \eqref{hfhe_ev2}.

\subsection{The HF/HE functional with positive and negative lotteries}\label{sec:PositiveNegative}

To rely on the HF/HE functional $H_{\lambda}$ in portfolio selection, we need to reformulate it in the context of mixed lotteries, namely when $Y \in \mathbb{R}$.
Suffice it to repeat the rationale behind \eqref{hfhe_ev1} and \eqref{hfhe_ev2} for a random variable $Y$, which admits both positive and negative values.

\noindent
Let $Y_{+}=\max\{Y,0\}$ and $Y_{-}= \min\{Y,0\}$, then $Y=Y_{+} + Y_{-}$, 
$|Y|=Y_{+} - Y_{-}$, and $\mu=\mathbb{E}[Y]=\mu_{+} + \mu_{-}$, where $\mu_{+}=\mathbb{E}[Y_{+}]$ and $\mu_{-}=\mathbb{E}[Y_{-}]$.
\noindent
Starting from \eqref{hfhe_ev1}, we can write
\begin{equation}
	\begin{split}
		H_1[Y] & =\mathbb{E}[Y-\mu] + \mu = \mathbb{E}[Y_+ + Y_- - \mu_+ - \mu_- ] +\mu = \\
		& = \mu + \mathbb{E}[(Y_+ - \mu_+)] + \mathbb{E}[(Y_- - \mu_-)].
	\end{split}
	\label{hfhe_1}
\end{equation}
Considering the relations
\begin{equation}
	\begin{split}
		Y_+ - \mu_+ &= (Y_+ - \mu_+)_+ + (Y_+ - \mu_+)_- \\
		Y_- - \mu_- &= (Y_- - \mu_-)_+ + (Y_- - \mu_-)_- \, , \\
	\end{split}
	\label{hfhe_2}
\end{equation}
and substituting them in \eqref{hfhe_1}, we have
\begin{equation}
	\begin{split}
		H_1[Y]&= \mu + \bigl \{ \mathbb{E} \bigl [(Y_+ - \mu_+)_+ + (Y_+ - \mu_+ )_- \bigr ] + \mathbb{E} \bigl [(Y_- - \mu_-)_+ + (Y_- - \mu_- )_- \bigr ] \bigr \} = \\
		& = \mu + \bigl \{ \mathbb{E} \bigl[ (Y_+ - \mu_+)_+ \bigr] + \mathbb{E} \bigl[ (Y_+ - \mu_+)_- \bigr]  \bigr \} + \bigl \{ \mathbb{E} \bigl[ (Y_- - \mu_-)_+ \bigr] + \mathbb{E} \bigl[ (Y_- - \mu_-)_- \bigr]  \bigr \} = \\
		&= \mu + 2\Bigl \{ \frac{1}{2}\mathbb{E} [ (Y_+ - \mu_+)_+ ] + \frac{1}{2}\mathbb{E} [ (Y_+ - \mu_+)_- ]  \Bigr \} + 2 \Bigl \{ \frac{1}{2}\mathbb{E} [ (Y_- - \mu_-)_+ ] + \frac{1}{2}\mathbb{E} [ (Y_- - \mu_-)_- ]  \Bigr \}. 
	\end{split}
	\label{hfhe_3}
\end{equation}
Similar to \eqref{hfhe_lambda}, the functional $H_1[Y]$ can be generalized as follows
\begin{equation}\label{hfhe_4}
H_2[Y] =\mu + 2\Bigl \{ \lambda_+ \mathbb{E} [ (Y_+ - \mu_+)_+ ] + (1-\lambda_+) \mathbb{E} [ (Y_+ - \mu_+)_- ]  \Bigr \} + 
		2 \Bigl \{ \lambda_- \mathbb{E} [ (Y_- - \mu_-)_+ ] + (1-\lambda_-)  \mathbb{E} [ (Y_- - \mu_-)_- ]  \Bigr \},    
\end{equation}
where $\lambda_+$ and $\lambda_-$ are the weights that the DM assigns to the outcomes above the mean in the case of gains and losses, respectively; $1-\lambda_+$ and $1-\lambda_-$ are the weights when the lotteries are below the mean in the case of gains and losses, respectively.

Using relations \eqref{hfhe_2}, $H_2[Y]$ can be expressed as
\begin{equation}
	\begin{split}
		H_2[Y] &=\mu + 2\Bigl \{ \lambda_+ \mathbb{E} [ (Y_+ - \mu_+)_+ ] - (1-\lambda_+) \mathbb{E} [ (Y_+ - \mu_+)_+ ]  \Bigr \} + \\
		&+ 2\Bigl \{ \lambda_- \mathbb{E} [ (Y_- - \mu_-)_+ ] - (1-\lambda_-)  \mathbb{E} [ (Y_- - \mu_-)_+]  \Bigr \} =\\
		& = \mu +2 \Bigl \{ (2\lambda_+ -1) \mathbb{E}[(Y_+-\mu_+)_+] \Bigr \} + 2 \Bigl \{ (2\lambda_- - 1 ) \mathbb{E}[(Y_- - \mu_-)_+] \Bigr \}.
		\label{hfhe_5}
	\end{split}
\end{equation}
Again from  \eqref{hfhe_2}, we have $\mathbb{E} \bigl [\abs{Y_+ - \mu_+}\bigr] = 2 \mathbb{E} [ (Y_+ - \mu_+)_+ ]$ and $\mathbb{E} \bigl [\abs{Y_- - \mu_-}\bigr] = 2 \mathbb{E} [ (Y_- - \mu_-)_+ ]$.
Thus, 
\begin{equation}
	\begin{split}
		H_2(Y) = \mu + (2\lambda_+ -1) \mathbb{E} \bigl [\abs{Y_+ - \mu_+}\bigr] + (2\lambda_- -1) \mathbb{E} \bigl [\abs{Y_- - \mu_-}\bigr]. 
		\label{hfhe_5_B}
	\end{split}
\end{equation}
We point out that the HF/HE functional \eqref{hfhe_5_B} depends on two parameters only, which, as shown in the next section, have a clear and well-established financial interpretation.

\subsubsection{Financial interpretation of the HF/HE parameters}
\label{sec:FinancialInterpretation}

If a DM adopts the value function $h(Y)$ to evaluate the lottery $Y$, then the certainty equivalent of $Y$, $CE_{Y}$, is defined as $h(CE_{Y})=h(Y)$: the DM is risk-adverse (risk-seeking) when $CE_{Y}<\mu$ ($CE_{Y}>\mu$), while for $CE_{Y}=\mu$, the DM is risk-neutral. 

In case $h(Y)=H_2(Y)$, we have $H_2(CE_{Y})=CE_{Y}$, hence $CE_{Y}=H_2(Y)$. Thus, accordingly, a DM is risk-adverse (risk-seeking) when $H_2(Y) < \mu$ ($H_2(Y) > \mu$), while if $H_2(Y) = \mu$, the DM is risk-neutral.
%
%

For positive lotteries ($Y>0$), \eqref{hfhe_5_B} becomes $H_2(Y)=\mu + (2\lambda_+ -1) \mathbb{E} \bigl [\abs{Y_+ - \mu_+}\bigr]$.
Therefore, a DM is 
\begin{itemize}
    \item risk-adverse, when $2\lambda_+ -1 < 0 \Leftrightarrow \lambda_+ \in [0, \frac{1}{2}) $;
    \item risk-seeking, when $2\lambda_+ -1 > 0 \Leftrightarrow \lambda_+ \in (\frac{1}{2},1] $;
    \item risk-neutral, when $2\lambda_+ -1 = 0 \Leftrightarrow \lambda_+ = \frac{1}{2} $.
\end{itemize}
For negative lotteries ($Y<0$), we have $H_2(Y)=\mu + (2\lambda_- -1) \mathbb{E} \bigl [\abs{Y_- - \mu_-}\bigr]$.
Hence, a DM is 
\begin{itemize}
    \item risk-adverse, when $2\lambda_- -1 < 0 \Leftrightarrow \lambda_- \in [0, \frac{1}{2}) $;
    \item risk-seeking, when $2\lambda_- -1 > 0 \Leftrightarrow \lambda_- \in (\frac{1}{2},1] $;
    \item risk-neutral, when $2\lambda_- -1 = 0 \Leftrightarrow \lambda_- = \frac{1}{2} $.
\end{itemize}
In the general case of mixed lotteries (i.e. $Y \in \mathbb{R}$), the DM is risk-averse when $H_2(Y) < \mu$, thus, if
\begin{equation*}
 (2\lambda_+ -1) \mathbb{E} \bigl [\abs{Y_+ - \mu_+}\bigr] + (2\lambda_- -1) \mathbb{E} \bigl [\abs{Y_- - \mu_-}\bigr] <  0.
\end{equation*}
Hence,$(2\lambda_+ -1) < (1-2\lambda_{-} ) \frac{\mathbb{E} \bigl [\abs{Y_- - \mu_-}\bigr]}{\mathbb{E} \bigl [\abs{Y_+ - \mu_+}\bigr]}$ yields
\begin{equation}\label{eq:interpr}
\lambda_+ < \frac{1}{2} + (\frac{1}{2} - \lambda_{-} ) \frac{\mathbb{E} \bigl [\abs{Y_- - \mu_-}\bigr]}{\mathbb{E} \bigl [\abs{Y_+ - \mu_+}\bigr]}.  
\end{equation}
Inequality \eqref{eq:interpr} can be rewritten as
\begin{equation}\label{eq:interpr_2}
 \lambda_{+} < -m \lambda_{-} + k \, ,   
\end{equation}
where $m = \displaystyle\frac{\mathbb{E} \bigl [\abs{Y_- - \mu_-}\bigr]}{\mathbb{E} \bigl [\abs{Y_+ - \mu_+}\bigr]} >0$ and $k=\frac{1}{2} (1+m) > \frac{1}{2}$.
\begin{figure}[tbp]
\centering
\includegraphics[width=0.5\textwidth]{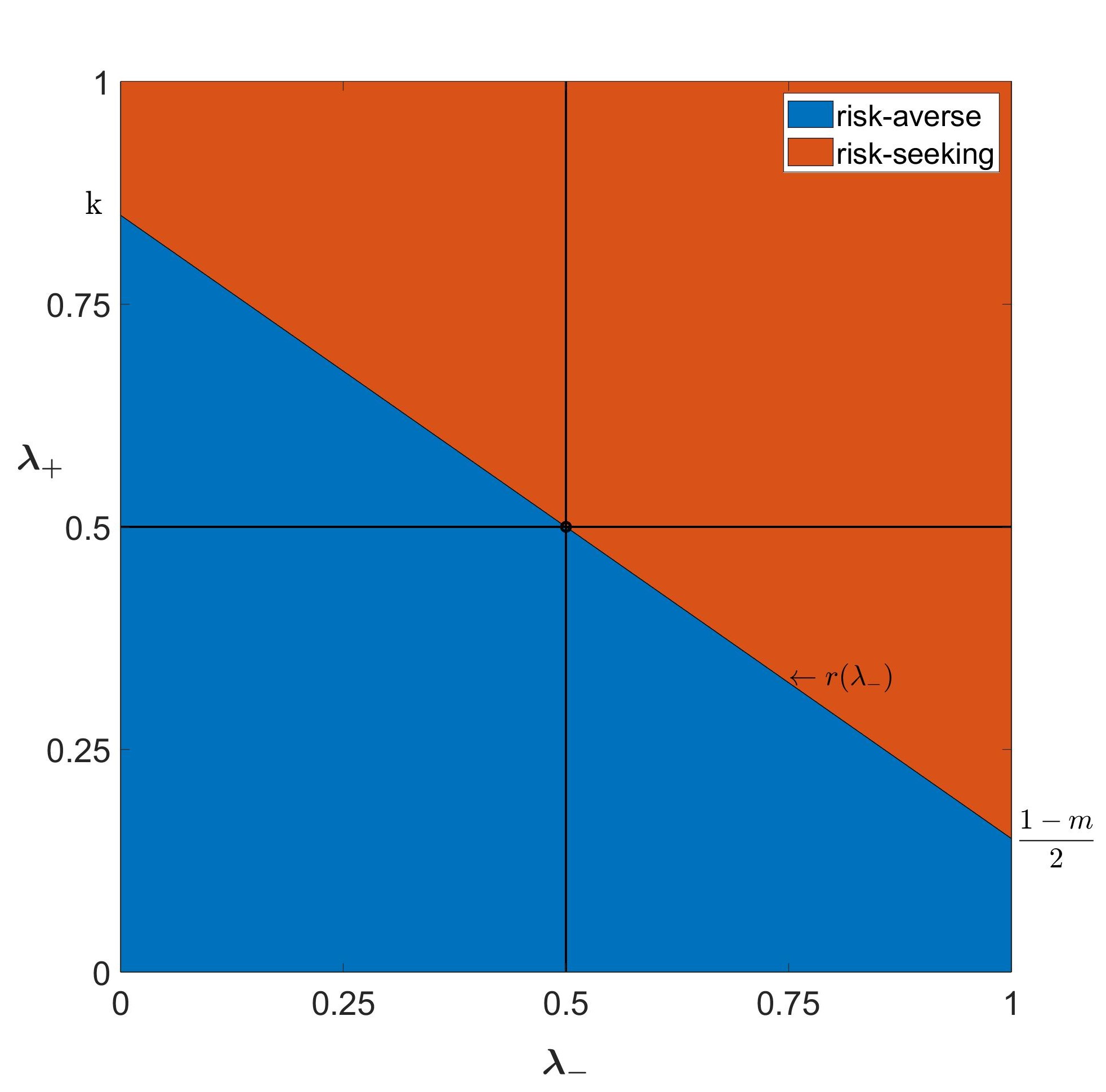} %
\caption{DM risk attitudes vs. $(\lambda_+, \lambda_- )$ for $m \in (0,1)$}
\label{fig:a}
\end{figure}

In Figure \ref{fig:a}, we report the typologies of risk attitudes of a DM, based on the values assumed by $\lambda_+ \in [0,1]$ and $\lambda_- \in [0,1]$ in the general case of mixed lotteries, with $m \in (0,1)$. 
The blue and red highlighted parts represent the values of $\lambda_+$ and $\lambda_-$ for which a DM is risk-averse and risk-seeking, respectively.
The values of $\lambda_+$ and $\lambda_-$ belonging to the line $r(\lambda_-): \lambda_{+} = -m \lambda_{-} + k$ represents a risk-neutral DM .

\begin{figure}[tbp]
\centering
\includegraphics[width=0.45\textwidth]{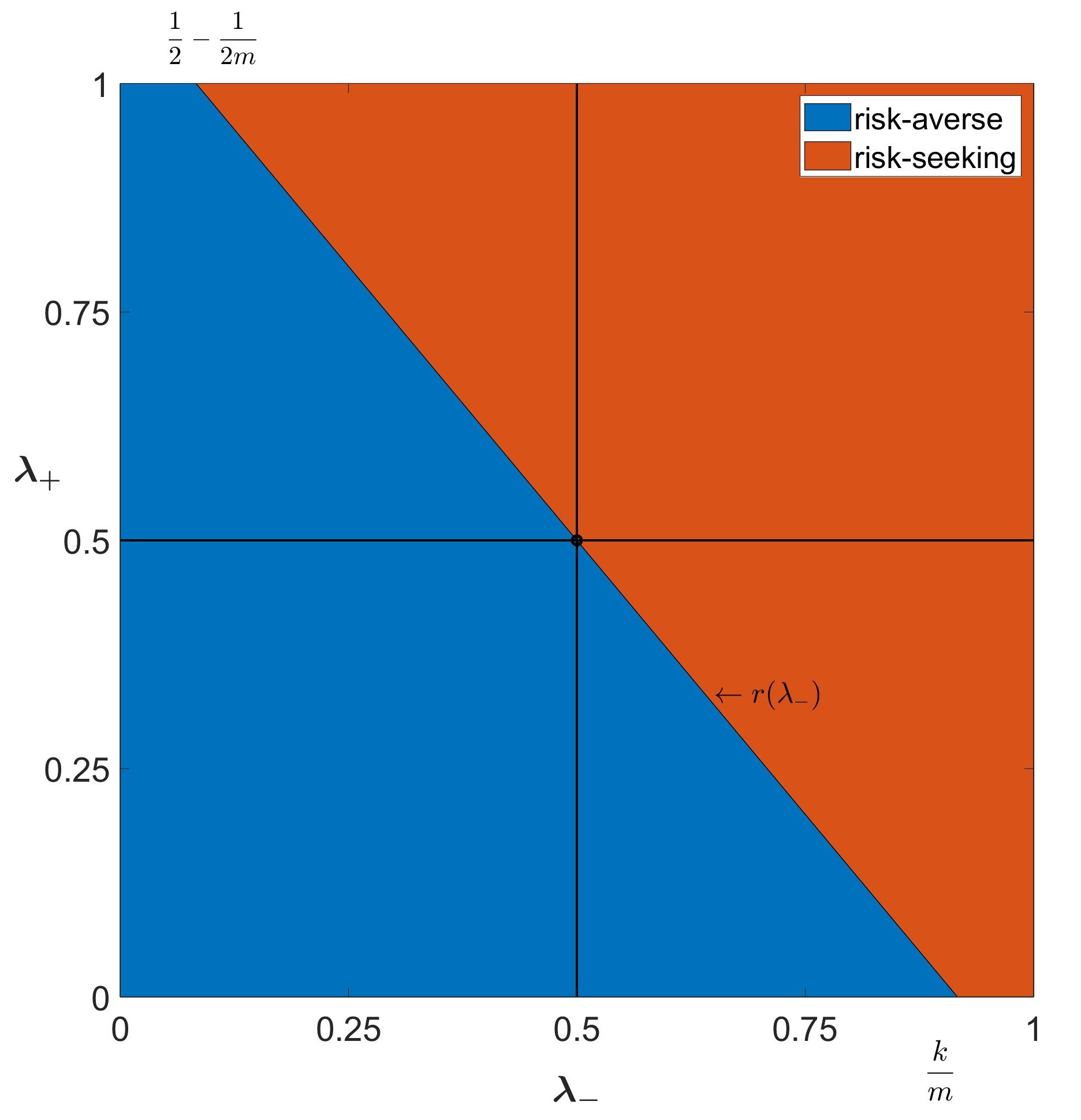}
\includegraphics[width=0.45\textwidth]{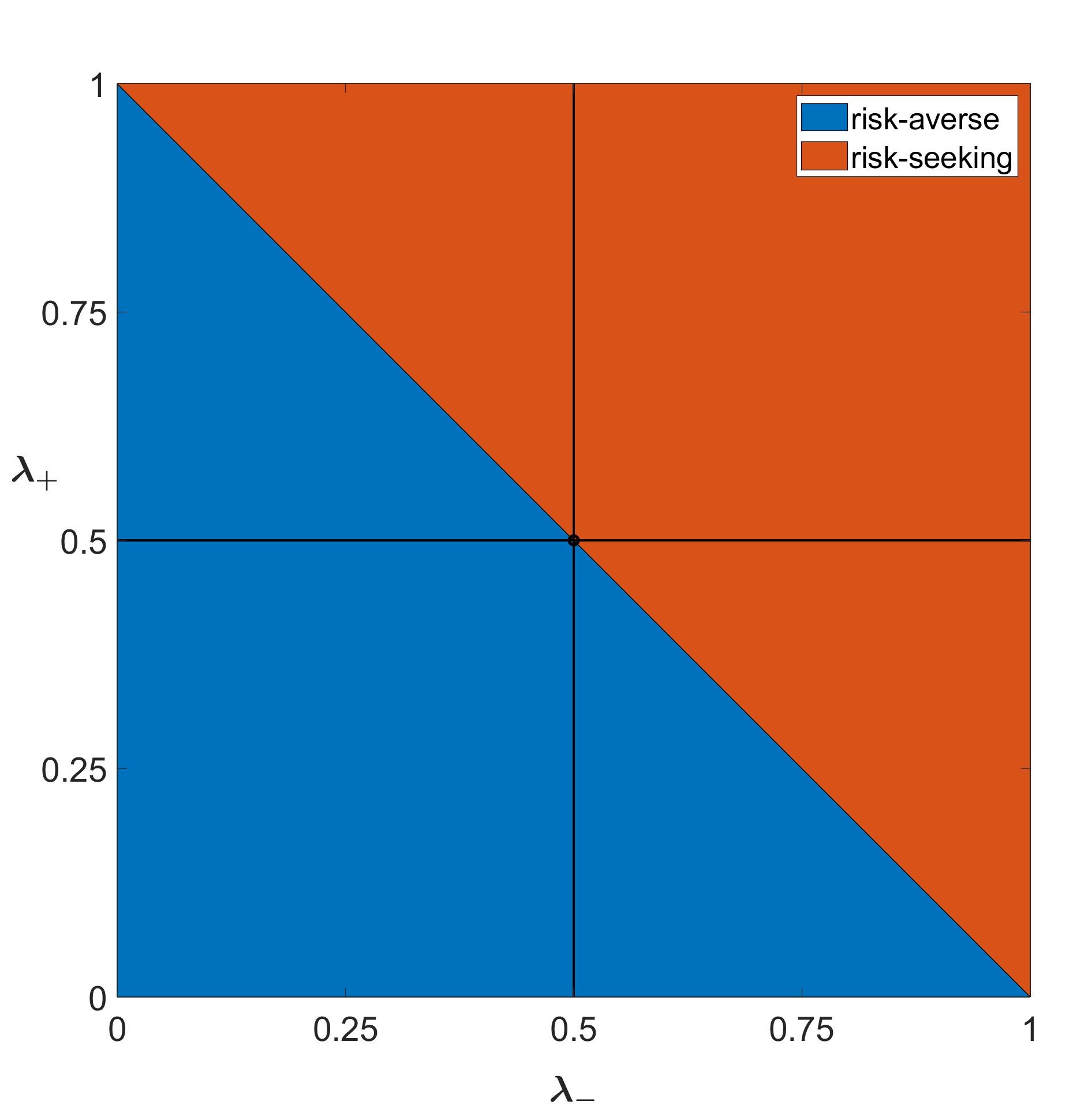}
\caption{DM risk attitudes vs. $(\lambda_+, \lambda_- )$ for $m > 1$ (left side) and $m = 1$ (right side)}
\label{fig:bc}
\end{figure}

\noindent
Figure \ref{fig:bc} shows the categories of risk attitudes of a  DM w.r.t. the values of $\lambda_+$ and $\lambda_-$ when $m > 1$ (left side) and $m = 1$ (right side).

\subsubsection{About the distortion of probabilities}
\label{sec:distortion}

A refinement of the functional defined in \eqref{hfhe_5_B} can be obtained by including the distortion of the lottery’s probabilities \citep[see, e.g.,][]{fehr2012probability,kahnemanandtversky1979,tversky1995risk}.
In the case of a discrete random variable $Y=\{y_1, y_2, \dots, y_T\}$ with probabilities $p=\{p_1, p_2, \dots, p_T\}$, it consists of transforming the probabilities $p_i$ by an appropriate weighting function. An example of weighting function is given by
\begin{equation}\label{eq:weighting}
 w(p_i,p,q)= \frac{p_i^q}{\sum_{j=1}^T p_j^q}  \, , 
\end{equation}
where $q$ is a suitable positive parameter that needs to be calibrated.

\begin{figure} [h!]
	\centering
 \subfloat[\emph{$q=0.7$}] {\includegraphics[width=.45\textwidth]{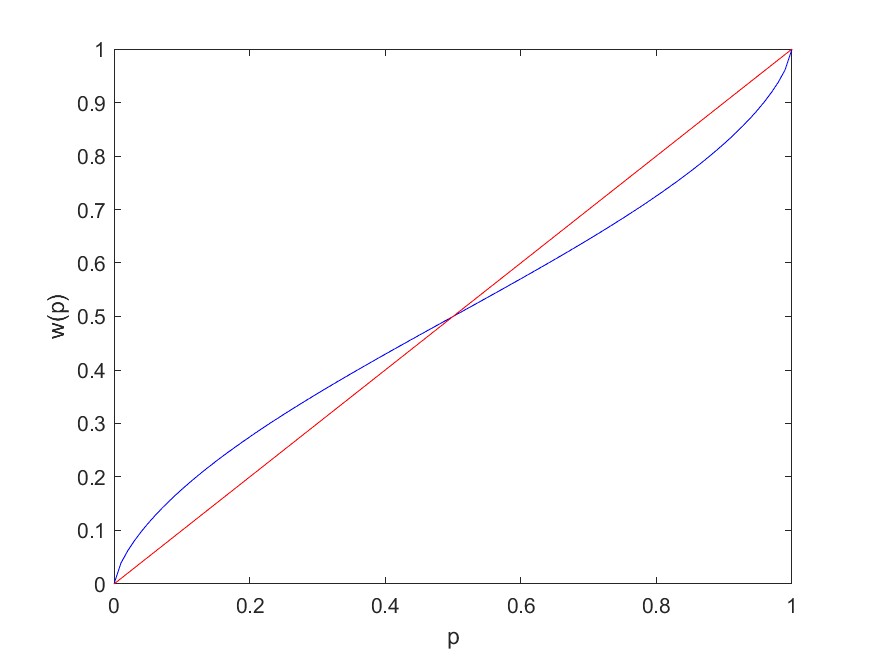}} \quad
 \subfloat[\emph{$q=1.4$}]
 {\includegraphics[width=.45\textwidth]{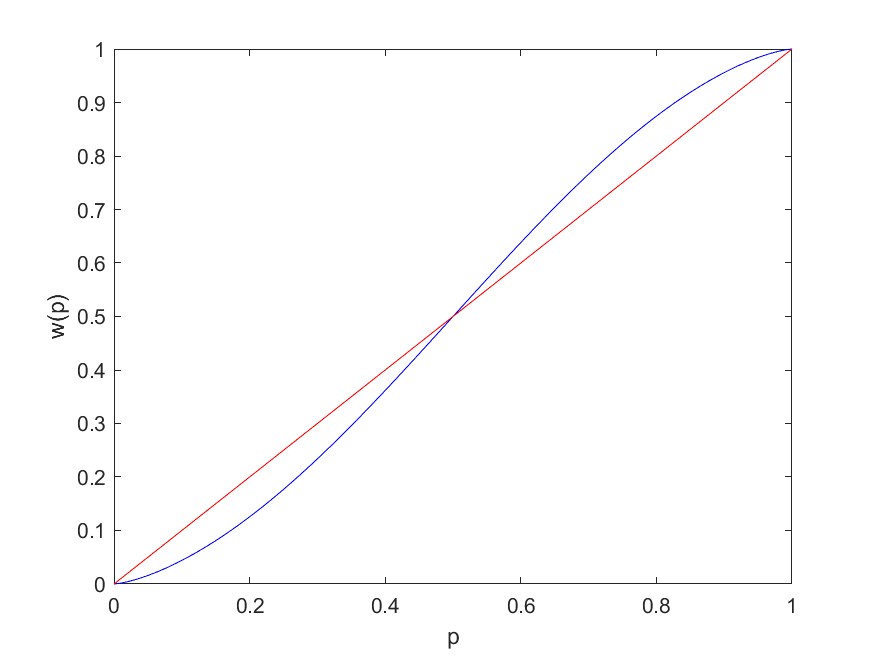}}
 \caption{Graph of the weighting function \eqref{eq:weighting}.}
 \label{fig:w_c}
\end{figure}

In Fig. \ref{fig:w_c}, we exhibit the behavior of the weighting function \eqref{eq:weighting} for a binary lottery $Y=\{(y_1,p); (y_2,1-p)\}$ in the case when $0<q<1$ e $q>1$.

\noindent
If $0<q<1$ (left side), the probability of the most likely events is underestimated, and that of the least likely events is overestimated; \emph{vice versa}, in case $q>1$ (right side).
When $q=1$, there is no probability distortion (see the red lines in Fig. \ref{fig:w_c}).
Then, expression \eqref{hfhe_5_B} can be generalized as follows:
\begin{equation}\label{eq:H_}
  H[Y] = \mu_q + (2\lambda_+ -1) \mathbb{E}_q \bigl [\abs{Y_+ - \mu_+}\bigr] + (2\lambda_- -1) \mathbb{E}_q \bigl [\abs{Y_- - \mu_-}\bigr], 
\end{equation}
where $\mu_q=\mathbb{E}_q[Y] = \sum_{j=1}^T w_q (p_j) y_j$, $\mathbb{E}_q [g(Y)]=\sum_{j=1}^T w_q (p_j) g(y_j)$, and $w_q(p_j)= \frac{p_j^q}{\sum_{i=1}^T p_i^q}$.
%


\subsection{The HF/HE portfolio selection model}\label{sec:HF/HEportfolioModel}

In this section, we adapt the HF/HE approach to portfolio selection problems.
We consider asset returns defined on a discrete probability space 
with $T$ states of nature, where each state has a probability of occurrence equal to $\pi_{t}$, $t=1, \ldots, T$.
Furthermore, we use a look-back approach where the realizations of the discrete random returns correspond to the historical scenarios, which are equally likely, i.e., $\pi_{t}=\frac{1}{T}$ $\forall t$, as typically assumed in portfolio optimization \citep[see, e.g.,][and references therein]{mansini2007conditional,carleo2017approximating,bellini2021risk}.
If $p_{kt}$ is the price of asset $k$ at time $t$, the return of asset $k$ at time $t$ is denoted by $r_{kt}=\frac{p_{kt}-p_{k(t-1)}}{p_{k(t-1)}}$.
Assuming an investment universe composed of $n$ assets and denoting by $x$ the vector of portfolio weights, the portfolio return at time $t$ is given by $R_t(x)=\sum_{k=1}^{n}x_k r_{kt}$.
For all models we consider, feasible portfolios are those identified by the budget constraint $\sum_{k=1}^n x_k =1$ and the no-short-selling requirements $x_k \ge 0 \quad k=1,\dots,n$.

\noindent
Note that in case of equally likely scenarios, the weighting function \eqref{eq:weighting} is $ w(\pi_{t},\pi,q) = \pi_{t} \equiv \frac{1}{T} $, $t=1, \ldots, T$, for every $q$.

\noindent
Thus, the HF/HE functional \eqref{hfhe_5_B} applied to the portfolio return becomes
\begin{equation}
		H[R(x)] = \mu(x)  + (2\lambda_+ -1) \frac{1}{T} \sum_{t=1}^{T} \abs{R_{t}^{+}(x)  - \mu^{+}(x)} + (2\lambda_- -1) \frac{1}{T} \sum_{t=1}^{T} \abs{R_{t}^{-}(x)  - \mu^{-}(x)},
		\label{hfhe_portfolio}
\end{equation}
where
\begin{gather}
	\begin{split}
        R(x) &= R^{+}(x) + R^{-}(x) \notag \\
        \mu(x) & = \sum_{k=1}^{n} x_k \mu_k \, , \qquad \mu_k = \frac{1}{T} \sum_{t=1}^T r_{kt}  \notag \\
		R_{t}^{+}(x) & = \max \Bigl( \sum_{k=1}^{n}x_k r_{kt} ,0 \Bigr) \notag \\
		\mu^{+}(x) & = \frac{1}{T} \sum_{t=1}^T R_{t}^{+}(x) = \frac{1}{T} \sum_{t=1}^T \max \Bigl( \sum_{k=1}^{n}x_k r_{kt} ,0 \Bigr) \notag \\
        R_{t}^{-}(x) & = \min \Bigl( \sum_{k=1}^{n}x_k r_{kt} ,0 \Bigr)  \notag \\
        \mu^{-}(x) & = \frac{1}{T} \sum_{t=1}^T R_{t}^{-}(x) = \frac{1}{T} \sum_{t=1}^T \min \Bigl( \sum_{k=1}^{n}x_k r_{kt} ,0 \Bigr). \notag 
	\end{split}
\end{gather}
Therefore, we obtain the following portfolio optimization problem:
\begin{equation}
	\begin{array}{cl}
		\underset{x}{\mbox{maximize}} & \displaystyle \sum_{k=1}^{n} x_k \mu_k +(2\lambda_+-1) \frac{1}{T} \sum_{t=1}^T \abs{ \max \bigl( \sum_{k=1}^n x_k r_{kt} ,0 \bigr)- \frac{1}{T} \sum_{t=1}^T \max \bigl( \sum_{k=1}^n x_k r_{kt},0 \bigr)}+ \\[5pt]
		& \displaystyle + (2\lambda_- - 1) \frac{1}{T} \sum_{t=1}^T \abs{\min \bigl( \sum_{k=1}^n x_k r_{kt} ,0 \bigr)- \frac{1}{T} \sum_{t=1}^T \min \bigl( \sum_{k=1}^n x_k r_{kt},0 \bigr)} \\[8pt]
		\mbox{s.t.} & \displaystyle \sum_{k=1}^n x_k =1\\[5pt]
		\label{hfhe_fmincon}
		& x_k \ge 0  \quad \qquad  k=1,\dots,n,
	\end{array}
\end{equation}
which is a nonconvex nonsmooth optimization problem.
However, as shown in the next section, it can be equivalently formulated as a MILP problem.

\noindent
Furthermore, under specific (but unrealistic) assumptions, optimal solutions of \eqref{hfhe_fmincon} can coincide with efficient portfolios obtained by relying on the Mean-MAD model developed by \cite{konnoandYamazaki1991} (see Remark \ref{rem:Mean-MAD}).
%

\subsection{A MILP reformulation of the HF/HE model}\label{sec:MILP}

The nonconvex nonsmooth optimization problem \eqref{hfhe_fmincon} can be equivalently recast according to a suitable MILP formulation, as shown below. Relying on such an alternative version of the model opens the door to different possible numerical treatments in order to address the problem in practice.

Preliminarily, one can introduce: 
(\textit{i}) $T$ auxiliary variables $d_{t}^{+}$ defined as the deviation of the portfolio return $\sum_{i=k}^{n} x_{k} r_{kt}$ from $0$ when $\sum_{k=1}^{n} x_{k} r_{kt} > 0$, and $0$ otherwise;
(\textit{ii}) $T$ auxiliary variables $d_{t}^{-}$ defined as the deviation of the portfolio loss $-\sum_{k=1}^{n} x_{k} r_{kt}$ from $0$ when $-\sum_{k=1}^{n} x_{k} r_{kt} > 0$, and $0$ otherwise.

We remark that the sets 
\[
\begin{array}{rcl}
{\mathcal{F}} \triangleq \Bigg\{(x, d^+, d^-) \in \mathbb R^n \times \mathbb R^T \times \mathbb R^T & : & \displaystyle d_t^+ = \max\left(\sum_{k=1}^n x_k r_{kt}, 0\right),\, d_t^- = \max\left(-\sum_{k=1}^n x_k r_{kt}, 0\right),\\[5pt]
& &\, t =1, \ldots, T, \, x \ge 0, \, \sum_{k=1}^n x_k = 1\Bigg\},\\[10pt]
{\mathcal{S}} \triangleq \Bigg\{(x, d^+, d^-) \in \mathbb R^n \times \mathbb R^T \times \mathbb R^T & : & \displaystyle d_t^+ - d_t^- = \sum_{k=1}^n x_k r_{kt},\, d_t^+ d_t^- = 0, \, d_t^+, d_t^- \ge 0, \, t=1, \ldots, T,\\[5pt]
& & x \ge 0, \, \sum_{k=1}^n x_k = 1\Bigg\}
\end{array}
\]
are equal. The inclusion ${\mathcal{F}} \subseteq {\mathcal{S}}$ follows easily from the definitions. As for ${\mathcal{S}} \subseteq {\mathcal{F}}$, suffice it to observe that, if $(y,b^+,b^-) \in {\mathcal{S}}$, $b_t^+ > 0$ yields $b_t^- = 0$ and, in turn, $b_t^+ = \sum_{k=1}^n y_k r_{kt}$, while $b_t^+ = 0$ entails $b_t^- = -\sum_{k=1}^n y_k r_{kt}$. Hence, the HF/HE model \eqref{hfhe_fmincon} can be equivalently recast as follows:
%
\begin{equation}  \label{eq:hfhe_}
\begin{array}{cll}
\displaystyle \underset{(x, d^{+}, d^{-})}{\mbox{minimize}} & \displaystyle -\sum_{k=1}^{n} x_k \mu_k -(2\lambda_+-1) \frac{1}{T} \sum_{t=1}^T \abs{d_{t}^{+} - \frac{1}{T} \sum_{t=1}^T d_{t}^{+}} - (2\lambda_- - 1) \frac{1}{T} \sum_{t=1}^T \abs{d_{t}^{-} - \frac{1}{T} \sum_{t=1}^T d_{t}^{-}} &  \\
\mbox{s.t.} & d_{t}^{+} - d_{t}^{-} = \sum_{k=1}^{n} x_k r_{kt} & \hspace*{-60pt} t=1,\ldots,T  \\
& d_{t}^{+}  d_{t}^{-} = 0  & \hspace*{-60pt} t=1,\ldots,T  \\
& \sum\limits_{k=1}^n x_k = 1  &  \\
& x_k\geq 0 & \hspace*{-60pt} k=1,\ldots,n \\
& d_{t}^{+}, d_{t}^{-} \geq 0 & \hspace*{-60pt} t=1,\ldots,T. 
\end{array}%
\end{equation}
As shown in Section \ref{sec:FinancialInterpretation}, considering a DM that is risk-adverse for gains and risk-seeking for losses, we have $(2\lambda_+-1) < 0$ and $(2\lambda_- -1) > 0$. 
In this case, we notice that $-(2\lambda_+-1) \frac{1}{T} \sum_{t=1}^T \abs{d_{t}^{+} - \frac{1}{T} \sum_{t=1}^T d_{t}^{+}}$ is a convex function, while $-(2\lambda_- - 1) \frac{1}{T} \sum_{t=1}^T \abs{d_{t}^{-} - \frac{1}{T} \sum_{t=1}^T d_{t}^{-}}$ is concave. Via the epigraph form of Problem \eqref{eq:hfhe_} \citep[see, e.g.,][Section 2.5]{grippo2023}, we get the following equivalent version of it:
\begin{equation}  \label{eq:hfhe_def_inter}
\begin{array}{cll}
\displaystyle \underset{(x, d^{+}, d^{-}, z^{+}, z^{-})}{\mbox{minimize}} & \displaystyle -\sum_{k=1}^{n} x_k \mu_k + (1 - 2\lambda_+) \frac{1}{T} \sum_{t=1}^T z_{t}^{+}+  (2\lambda_- - 1) \frac{1}{T} \sum_{t=1}^T z_{t}^{-} &  \\
\mbox{s.t.} & - z_{t}^{+} - (d_{t}^{+} - \frac{1}{T} \sum_{t=1}^T d_{t}^{+})  \leq  0 & t=1,\ldots,T  \\
 & - z_{t}^{+} + (d_{t}^{+} - \frac{1}{T} \sum_{t=1}^T d_{t}^{+}) \leq  0 & t=1,\ldots,T  \\
 & - z_{t}^{-} - |d_t^- - \frac{1}{T} d_t^-|\leq 0 & t=1,\ldots,T  \\
& d_{t}^{+} - d_{t}^{-} = \sum_{k=1}^{n} x_{k} r_{kt} & t=1,\ldots,T  \\
& d_{t}^{+}  d_{t}^{-} = 0  & t=1,\ldots,T  \\
& \sum\limits_{k=1}^n x_k = 1  &  \\
& x_k\geq 0 & k=1,\ldots,n \\
& d_{t}^{+}, d_{t}^{-} \geq 0 & t=1,\ldots,T. 
\end{array}%
\end{equation}
We finally deal with the difficult-to-handle disjunctive-type constraints $- z_{t}^{-} - |d_t^- - \frac{1}{T} d_t^-|\leq 0$ and complementarity conditions $d_{t}^{+} d_{t}^{-} = 0$, $t=1, \ldots, T$, in \eqref{eq:hfhe_def_inter} through the big-M reformulation, thus obtaining the mixed-integer linear program
\begin{equation}  \label{eq:hfhe_def}
\begin{array}{cll}
\displaystyle \underset{(x, d^{+}, d^{-}, z^{+}, z^{-}, \bar{y}, \tilde{y}, w)}{\mbox{minimize}} & \displaystyle -\sum_{k=1}^{n} x_k \mu_k + (1 - 2\lambda_+) \frac{1}{T} \sum_{t=1}^T z_{t}^{+}+  (2\lambda_- - 1) \frac{1}{T} \sum_{t=1}^T z_{t}^{-} &  \\
\mbox{s.t.} & - z_{t}^{+} - (d_{t}^{+} - \frac{1}{T} \sum_{t=1}^T d_{t}^{+})  \leq  0 & t=1,\ldots,T  \\
 & - z_{t}^{+} + (d_{t}^{+} - \frac{1}{T} \sum_{t=1}^T d_{t}^{+}) \leq  0 & t=1,\ldots,T  \\
 & - z_{t}^{-} \leq M (1- \bar{y}_t) & t=1,\ldots,T  \\
 & - z_{t}^{-} - (d_{t}^{-} - \frac{1}{T} \sum_{t=1}^T d_{t}^{-})  \leq  M (1- \bar{y}_t) & t=1,\ldots,T  \\
 & z_{t}^{-} \leq  M (1- \tilde{y}_t) & t=1,\ldots,T  \\
 & - z_{t}^{-} + (d_{t}^{-} - \frac{1}{T} \sum_{t=1}^T d_{t}^{-}) \leq  M (1- \tilde{y}_t) & t=1,\ldots,T  \\
 & \bar{y}_t, \tilde{y}_t \in \{0, 1\} & t=1,\ldots,T  \\
 & \bar{y}_t + \tilde{y}_t = 1 &  t=1,\ldots,T\\
& d_{t}^{+} - d_{t}^{-} = \sum_{k=1}^{n} x_{k}r_{kt} & t=1,\ldots,T  \\
& d_{t}^{+} \le M w_t  & t=1,\ldots,T  \\
& d_{t}^{-} \le M (1 - w_t)  & t=1,\ldots,T  \\
& w_t \in \{0,1\} & t=1,\ldots,T \\
& \sum\limits_{k=1}^n x_k = 1  &  \\
& x_k\geq 0 & k=1,\ldots,n \\
& d_{t}^{+}, d_{t}^{-} \geq 0 & t=1,\ldots,T, 
\end{array}%
\end{equation}
where $M$ is a suitably large positive scalar.

\section{Other portfolio selection models to be compared with}\label{sec:OtherModels}

We introduce several portfolio selection models to be compared with the HF/HE approach.
More precisely, in Section \ref{sec:PTmodel}, we briefly describe the behavioral portfolio selection model based on PT.
In Section \ref{sec:MinV-MAD}, we present two minimum risk approaches, using variance and MAD, as a risk measures, respectively.
Furthermore, we also test the so-called Equally Weighted (EW) portfolio, defined by $x_k=\frac{1}{n}$, $k=1, \cdots, n$, whose performances would seem to be barely improved by optimized portfolios in practice \citep[][]{demiguel2009optimal}.

\subsection{Prospect Theory portfolio}\label{sec:PTmodel}

\cite{kahnemanandtversky1979} suggest to interpret individuals' experimentally observed economic choices by modeling their satisfaction level through a nonconcave function. 
A DM exhibits a risk-averse attitude toward gains and a risk-seeking attitude toward losses. A crucial property of PT is loss aversion: individuals are more sensitive to losses than gains \citep[see][]{sun2009loss}.
Furthermore, among different alternatives, investors evaluate potential gains and losses with respect to a reference point. They overestimate the probabilities of significant gains and losses.
Accordingly, on the one hand, the satisfaction levels of a DM are modeled by a function that is concave for gains and convex for losses, thus incorporating a risk-averse attitude for gains and a risk-seeking attitude for losses.
The value function is steeper below the reference since people tend to focus more on losses than gains.
On the other hand, a DM transforms objective probabilities through a weighting function, by underestimating medium-high probabilities and overestimating low probabilities associated with extreme events (see Section \ref{sec:distortion}). 

To make comparisons, we refer to the PT-based behavioral portfolio selection model proposed by \cite{de2007computational}:
\begin{equation}\label{PT_fmincon}
\begin{array}{cl}
	\underset{x}{\mbox{maximize}} & \displaystyle \frac{1}{T} \sum_{t=1}^T \Bigl[ \Bigl( \Bigl (\sum_{k=1}^n x_k r_{kt} \Bigr )_+ \Bigr )^\alpha  - \beta \Bigl ( - \Bigl (\sum_{k=1}^n x_k r_{kt} \Bigr )_- \Bigr )^\alpha \Bigr]\\
\mbox{s.t.} & \displaystyle \sum_{k=1}^n x_k =1\\
& x_k \ge 0 \quad k=1,\dots,n,
\end{array}
\end{equation}
where $\alpha \le 1$ leads to a concave objective for gains and convex for losses, and controls the DM's risk attitude. Furthermore, $\beta$ represents loss aversion. Also, if $\beta > 1$, the function is asymmetric, so that DM is more loss-averse than gain-averse.
Model \eqref{PT_fmincon}, thus, is a nonconvex nonsmooth optimization problem. As in \cite{de2007computational}, we fix $\alpha=0.88$ and $\beta=2.25$.

\subsection{Minimum variance and minimum MAD portfolios} \label{sec:MinV-MAD}

The long-only minimum variance portfolio \citep[][]{Mark:52,Mark:59} can be obtained by solving the following convex quadratic programming problem:
\begin{equation}  \label{min_v}
\begin{array}{cl}
\underset{x}{\mbox{minimize}} & \displaystyle \sum_{k=1}^{n}\sum_{j=1}^{n}\sigma_{k,j}x_k x_j\\
\mbox{s.t.} & \sum\limits_{k=1}^n x_k = 1 \\
& x_k\geq 0 \quad k=1,\ldots,n,
\end{array}%
\end{equation}
where $\sigma_{k,j}$ is the covariance of returns of asset $k$ and asset $j$, $k, j=1,\dots n$.

An alternative deviation risk measure in the sense of \cite{rockafellar2006generalized}, which has been often used in the literature \citep[see, e.g.,][and references therein]{ararat2021mad,carleo2017approximating,cesarone2015linear,cesarone2016optimally,cesarone2020computational}, is the MAD: it is defined as the expected value of the absolute deviation of the portfolio return from its mean.
Thus, the long-only minimum MAD portfolio \citep{konnoandYamazaki1991} can be obtained by solving the following optimization problem:
\begin{equation}\label{min_mad}
\begin{array}{cl}
\underset{x}{\mbox{minimize}} & \mbox{MAD}(x) = \displaystyle \frac{1}{T} \sum_{t=1}^{T} \left\lvert\sum_{k=1}^{n} r_{k,t} x_{k} -\sum_{k=1}^{n} \mu_{k} x_{k}\right\rvert\\
\mbox{s.t.} & \sum\limits_{k=1}^n x_k = 1\\
& x_k\geq 0, \quad k=1,\ldots,n,
\end{array}%
\end{equation}
where $\mu_k$ represents the expected return of asset $k$.
As in \cite{konnoandYamazaki1991}, one can linearize problem \eqref{min_mad} by introducing $T$ auxiliary variables
$d_{t}$ defined as the absolute deviation of the portfolio return from its mean, thus obtaining, following a similar logic as in Section \ref{sec:MILP}, the Linear Programming (LP) problem 
%
\begin{equation}\label{min_mad_lin}
\begin{array}{cll}
\underset{(x,d)}{\mbox{minimize}} & \displaystyle \frac{1}{T} \sum_{t=1}^{T} d_{t} &\\
\mbox{s.t.} & \displaystyle d_{t} \geq \sum_{k=1}^{n} (r_{k,t} - \mu_k) x_{k} & t=1,\ldots,T\\
& \displaystyle d_{t} \geq -\sum_{k=1}^{n} (r_{k,t} - \mu_k) x_{k} & t=1,\ldots,T\\
& \sum\limits_{k=1}^n x_k = 1  &  \\
& x_k\geq 0 & k=1,\ldots,n.
\end{array}%
\end{equation}
\begin{remark}[HF/HE vs. Mean-MAD]\label{rem:Mean-MAD}
Assuming $R(x) >0$, if $\lambda_{+} \in [0, \frac{1}{2})$, the HF/HE portfolio coincides with the Mean-MAD portfolio
\begin{equation}
    x^{*}=\argmin \{- \mu(x) + (1-2 \lambda_{+}) MAD(x): \, x \in \Delta\}.
\end{equation}
Assuming $R(x) <0$, if $\lambda_{-} \in [0, \frac{1}{2})$, the HF/HE portfolio coincides with the Mean-MAD portfolio
\begin{equation}
    x^{*}=\argmin \{- \mu(x) + (1-2 \lambda_{-}) MAD(x): \, x \in \Delta\}.
\end{equation}
Whenever $R(x) \in \mathbb{R}$, Mean-MAD and HF/HE portfolios are different.
Indeed, the HF/HE model weighs positive and negative outcomes differently, while 
MAD is a symmetric risk measure and equally penalizes outcomes above and below the portfolio expected return.
\end{remark}

\section{Empirical analysis}\label{sec:EmpiricalAnalysis}

By means of a thorough empirical analysis, we test the performances of HF/HE portfolios in comparison with EW, minimum variance and MAD, and PT ones. 

Summing up, in Section \ref{sec:Description}, we describe datasets and performance measures. We, then, discuss the computational results: more precisely, in Section \ref{sec:ComputationalResults}, we test and validate HF/HE portfolios, choosing parameters $\lambda_+$ and $\lambda_-$ values that satisfy old and new paradoxes \citep[]{kahnemanandtversky1979, birnbaum2008new}; in Section \ref{sec:Flexibility}, we explore the flexibility of the HF/HE approach in terms of the out-of-sample performances by varying $\lambda_+$ and $\lambda_-$ outside the paradoxes' constraints.
%

For the out-of-sample performance analysis, we adopt a rolling time window mechanism: using a time window of 500 (daily) observations (in-sample period), we solve the problem for overlapping windows constructed by moving forward in time in steps of 20 days. The optimal portfolio that is computed in an in-sample period is then maintained for the next 20 days (out-of-sample period), so that we can evaluate its performances by considering the aforementioned measures.
%

All experiments have been implemented on a laptop with an Intel(R) Core(TM) i7-7500U CPU @ 2.70 GHz processor and 8 GB of RAM, using MATLAB R2022b. We address problems \eqref{hfhe_fmincon} and \eqref{PT_fmincon} using a multi-start heuristic, while the remaining models are coped with by means of standard built-in solvers.  

\subsection{Description of datasets and performance measures}\label{sec:Description}

We consider the following datasets that are publicly available on \url{https://www.francescocesarone.com/data-sets} and consist of daily prices, adjusted for dividends and stock splits and retrieved from Refinitiv \citep[see, for more details,][]{cesarone2022does}:
\begin{itemize}
    \item \textbf{DIJA} (Dow Jones Industrial Average, USA): composed of 28 assets and 3715 observations (October 2006 - December 2020);
    \item \textbf{NASDAQ 100} (National Association of Securities Dealers Automated Quotation, USA): composed of 54 assets and 3715 observations (October 2006 - December 2020);
    \item \textbf{FTSE 100} (Financial Times Stock Exchange, UK): composed of 80 assets and 3715 observations (October 2006 - December 2020).
\end{itemize}
%
To evaluate the out-of-sample performances of the portfolio strategies we analyze, we use the following measures that are typically resorted to in the literature \citep[see, e.g.,][and references therein]{cesarone2015linear,cesarone2016optimally,cesarone2020optimization,bruni2017exact,cesarone2017minimum}:
%

\begin{itemize}
\item out‐of‐sample daily expected return $\hat{\mu}^{out}$ (\textbf{ExpRet}) and out‐of‐sample daily volatility $\hat{\sigma}^{out}$ (\textbf{Vol}).
\item \textbf{Sharpe} ratio \citep[]{sharpe1966mutual, sharpe1998sharpe}, i.e. gain per unit of risk:
\begin{equation*}
\mbox{\bf{Sharpe}} =\frac{\hat{\mu}^{out} - r_f}{\hat{\sigma}^{out}} \, ,
\label{SR}
\end{equation*}
where we fix $r_f=0$;
\item Maximum Drawdown \citep[\textbf{MaxDD}, see, e.g.,][]{chekhlov2005drawdown}, i.e.
%
\begin{equation*}
\mbox{\bf{MaxDD}} = \min_{T^{in}+1 \leq t \leq T} DD_t, \quad  DD_t=\frac{W_t - \max\limits_{T^{in}+1 \leq \tau \leq t} W_\tau}{\max\limits_{T^{in}+1 \leq \tau \leq t} W_\tau}
 \qquad {t \in \{T^{in}+1,\dots,T}\},
\label{DD_t}
\end{equation*}
where $T^{in}$ is the length of the in-sample window, $W_{\tau }=W_{\tau -1}(1+R_{\tau }^{out})$ is the value of wealth after $\tau $
periods, $W_{T^{in}}=1$ is the initial wealth, and $R_{\tau}^{out}$ denotes the out-of-sample portfolio returns for each portfolio strategy;

\item \textbf{Sortino} ratio \citep{sortino2001managing}, which, again, is a gain-risk ratio performance measure defined as follows:
\begin{equation*}
\mbox{\bf{Sortino}}= \frac{\hat{\mu}^{out} - r_f}{\sqrt{\mathbb{E},[(r^{out}-r_f)_{-}^{2}]}} 
\label{SOR}
\end{equation*}
where ${\sqrt{\mathbb{E}[(r^{out}-r_f)_{-}^{2}]}}$ represents the Target Downside Deviation and $r_f=0$;
\item \textbf{Rachev} ratio \citep{biglova2004different}, which measures the upside potential and is defined as the ratio between expected return in the best $\alpha$\% values of $R^{out}-r_f$ and expected losses in the worst $\beta$\% values of $R^{out}-r_f$:
\begin{equation*}
\mbox{\bf{Rachev}}= \frac{\mbox{CVaR}_{\alpha}(r_f - {R}^{out})}{\mbox{CVaR}_{\beta}(R^{out}-r_f)}, 
\label{SOR_}
\end{equation*}
where $\alpha=\beta=5$ and $r_f=0$;
\item average Return On Investment \citep[\textbf{aveROI}, see, e.g.,][]{phillips2005return}, which indicates the return generated by each portfolio strategy for a given time horizon $\Delta \tau$:
\begin{equation*}
   ROI_{\tau} = \frac{W_{\tau}-W_{\tau-\Delta \tau}}{W_{\tau -\Delta \tau}} \qquad \tau = \Delta \tau + 1, \dots ,T,
\end{equation*}
where $W_{\tau - \Delta \tau}$ represents the capital invested at the beginning of the time horizon, $W_{\tau}=W_{\tau - \Delta \tau} \prod_{t=\tau-\Delta \tau +1}^\tau (1+R_{t}^{out})$ is the portfolio wealth, and $T$ is the entire number of the available realizations; 
\item Herfindahl index \citep{rhoades1993herfindahl}, which is a concentration measure, defined as $\mbox{HI}=\sum_{k=1}^{n} x_{k}^{2}$.
%
%
The normalized version of the index (\textbf{NHI}) is
\begin{equation*}
\mbox{\bf{NHI}}= \displaystyle\frac{1-HI}{1-\frac{1}{n}}.
\label{NHI}
\end{equation*} 
Hence, $\mbox{\bf{NHI}}=0$ represents the maximum concentration, while $\mbox{\bf{NHI}}=1$ indicates the maximum diversification;    
\item average number of assets (\textbf{ave \#}) selected by each portfolio strategy.
\end{itemize}


\subsection{Comparison of out-of-sample performance results}\label{sec:ComputationalResults}

We collect in some tables the results of our empirical analysis for all the portfolio strategies we examine. To facilitate the interpretation of the data, we exhibit the rank of the performance measures for the different approaches with different colors: for each row corresponding to a specific performance measure, the colors range from deep green to deep red, where deep green represents the best value and deep red is the worst.

In Tables \ref{1}, \ref{2}, and \ref{3}, we report the out-of-sample performance results for the five portfolio selection models, considering DIJA, NASDAQ100, and FTSE100 datasets, respectively.  
%
%

We point out that, preliminarily, we examine HF/HE portfolios setting parameters $\lambda_+$ and $\lambda_-$ within a range of values so that all the paradoxes described in \cite{kahnemanandtversky1979} and \cite{birnbaum2008new} are satisfied: thus, $\lambda_+ \in [0.29, 0.32]$ and $\lambda_- \in [0.68, 0.72]$, according to the results in \citep[]{cencietal2015,corradini2022half}. Then, among the portfolios that have been obtained for different values of the parameters, we select the best one, namely the portfolio for which $\lambda_+=0.30$ and $\lambda_-=0.69$.

\begin{table}[htbp]
  \centering
  \caption{Out-of-sample performance results for DIJA}
   \label{1}
   \scalebox{0.9}{
    \begin{tabular}{|l|rrrrr|}
    \toprule
    \multicolumn{1}{|c|}{\textbf{DIJA}} & \multicolumn{1}{c}{\textbf{EW}} & \multicolumn{1}{c}{\textbf{MinV}} & \multicolumn{1}{p{5.545em}}{\textbf{MinMAD}} & \multicolumn{1}{c}{\textbf{PT}} & \multicolumn{1}{p{5.545em}|}{\textbf{HF/HE 0.30-0.69}} \\
    \midrule
    \textbf{ExpRet} & \cellcolor[rgb]{ .792,  .863,  .506}0,048\% & \cellcolor[rgb]{ .973,  .412,  .42}0,026\% & \cellcolor[rgb]{ .988,  .729,  .478}0,032\% & \cellcolor[rgb]{ 1,  .922,  .518}0,035\% & \cellcolor[rgb]{ .388,  .745,  .482}0,073\% \\
    \textbf{Vol} & \cellcolor[rgb]{ .996,  .788,  .494}0,013 & \cellcolor[rgb]{ .388,  .745,  .482}0,010 & \cellcolor[rgb]{ .392,  .745,  .482}0,010 & \cellcolor[rgb]{ 1,  .922,  .518}0,011 & \cellcolor[rgb]{ .973,  .412,  .42}0,018 \\
    \textbf{Sharpe} & \cellcolor[rgb]{ .604,  .808,  .498}0,038 & \cellcolor[rgb]{ .973,  .412,  .42}0,027 & \cellcolor[rgb]{ 1,  .922,  .518}0,033 & \cellcolor[rgb]{ .996,  .878,  .506}0,033 & \cellcolor[rgb]{ .388,  .745,  .482}0,040 \\
    \textbf{MaxDD} & \cellcolor[rgb]{ .988,  .702,  .475}-0,466 & \cellcolor[rgb]{ .388,  .745,  .482}-0,332 & \cellcolor[rgb]{ .796,  .863,  .506}-0,366 & \cellcolor[rgb]{ 1,  .922,  .518}-0,383 & \cellcolor[rgb]{ .973,  .412,  .42}-0,580 \\
    \textbf{Sortino} & \cellcolor[rgb]{ .663,  .827,  .502}0,054 & \cellcolor[rgb]{ .973,  .412,  .42}0,038 & \cellcolor[rgb]{ 1,  .922,  .518}0,047 & \cellcolor[rgb]{ .996,  .867,  .506}0,046 & \cellcolor[rgb]{ .388,  .745,  .482}0,060 \\
    \textbf{Rachev} & \cellcolor[rgb]{ .843,  .878,  .51}0,968 & \cellcolor[rgb]{ .973,  .412,  .42}0,921 & \cellcolor[rgb]{ 1,  .922,  .518}0,946 & \cellcolor[rgb]{ .984,  .639,  .463}0,932 & \cellcolor[rgb]{ .388,  .745,  .482}1,030 \\
    \textbf{AveROI} & \cellcolor[rgb]{ .8,  .867,  .51}45\% & \cellcolor[rgb]{ .973,  .412,  .42}26\% & \cellcolor[rgb]{ .988,  .71,  .475}32\% & \cellcolor[rgb]{ 1,  .922,  .518}35\% & \cellcolor[rgb]{ .388,  .745,  .482}65\% \\
    \textbf{NHI} & \multicolumn{1}{c}{-} & \cellcolor[rgb]{ .996,  .918,  .514}0,818 & \cellcolor[rgb]{ .388,  .745,  .482}0,829 & \cellcolor[rgb]{ .933,  .902,  .514}0,820 & \cellcolor[rgb]{ .973,  .412,  .42}0,427 \\
    \textbf{ave \#} & 28    & 11,6  & 11,3  & 11,3  & 2,9 \\
    \bottomrule
    \end{tabular}%
    }%
\end{table}%
%

\begin{table}[htbp]
  \centering
  \caption{Out-of-sample performance results for NASDAQ 100}
  \label{2}
  \scalebox{0.9}{
    \begin{tabular}{|l|crrrr|}
    \toprule
    \multicolumn{1}{|c|}{\textbf{NASDAQ100}} & \textbf{EW} & \multicolumn{1}{c}{\textbf{MinV}} & \multicolumn{1}{p{5.545em}}{\textbf{MinMAD}} & \multicolumn{1}{c}{\textbf{PT}} & \multicolumn{1}{p{5.545em}|}{\textbf{HF/HE 0.30-0.69}} \\
    \midrule
    \textbf{ExpRet} & \multicolumn{1}{r}{\cellcolor[rgb]{ .898,  .894,  .514}0,068\%} & \cellcolor[rgb]{ .973,  .439,  .424}0,040\% & \cellcolor[rgb]{ .973,  .412,  .42}0,039\% & \cellcolor[rgb]{ 1,  .922,  .518}0,053\% & \cellcolor[rgb]{ .388,  .745,  .482}0,141\% \\
    \textbf{Vol} & \multicolumn{1}{r}{\cellcolor[rgb]{ 1,  .867,  .51}0,014} & \cellcolor[rgb]{ .404,  .749,  .482}0,010 & \cellcolor[rgb]{ .388,  .745,  .482}0,010 & \cellcolor[rgb]{ 1,  .922,  .518}0,012 & \cellcolor[rgb]{ .973,  .412,  .42}0,024 \\
    \textbf{Sharpe} & \multicolumn{1}{r}{\cellcolor[rgb]{ .733,  .847,  .506}0,049} & \cellcolor[rgb]{ .973,  .478,  .431}0,040 & \cellcolor[rgb]{ .973,  .412,  .42}0,039 & \cellcolor[rgb]{ 1,  .922,  .518}0,043 & \cellcolor[rgb]{ .388,  .745,  .482}0,058 \\
    \textbf{MaxDD} & \multicolumn{1}{r}{\cellcolor[rgb]{ .992,  .812,  .494}-0,524} & \cellcolor[rgb]{ .557,  .796,  .494}-0,348 & \cellcolor[rgb]{ .388,  .745,  .482}-0,326 & \cellcolor[rgb]{ 1,  .922,  .518}-0,424 & \cellcolor[rgb]{ .973,  .412,  .42}-0,582 \\
    \textbf{Sortino} & \multicolumn{1}{r}{\cellcolor[rgb]{ .776,  .859,  .506}0,070} & \cellcolor[rgb]{ .973,  .463,  .427}0,056 & \cellcolor[rgb]{ .973,  .412,  .42}0,056 & \cellcolor[rgb]{ 1,  .922,  .518}0,061 & \cellcolor[rgb]{ .388,  .745,  .482}0,086 \\
    \textbf{Rachev} & \multicolumn{1}{r}{\cellcolor[rgb]{ .941,  .906,  .518}0,949} & \cellcolor[rgb]{ .996,  .882,  .51}0,937 & \cellcolor[rgb]{ 1,  .922,  .518}0,937 & \cellcolor[rgb]{ .973,  .412,  .42}0,931 & \cellcolor[rgb]{ .388,  .745,  .482}1,054 \\
    \textbf{AveROI} & \multicolumn{1}{r}{\cellcolor[rgb]{ .992,  .922,  .518}68\%} & \cellcolor[rgb]{ .973,  .424,  .42}41\% & \cellcolor[rgb]{ .973,  .412,  .42}41\% & \cellcolor[rgb]{ 1,  .922,  .518}66\% & \cellcolor[rgb]{ .388,  .745,  .482}233\% \\
    \textbf{NHI} & -     & \cellcolor[rgb]{ .89,  .89,  .514}0,850 & \cellcolor[rgb]{ .388,  .745,  .482}0,854 & \cellcolor[rgb]{ .996,  .918,  .514}0,849 & \cellcolor[rgb]{ .973,  .412,  .42}0,448 \\
    \textbf{ave \#} & 54    & 14,6  & 14,5  & 16,6  & 3,6 \\
    \bottomrule
    \end{tabular}%
    }%
\end{table}%
%

\begin{table}[htbp]
  \centering
  \caption{Out-of-sample performance results for FTSE 100}
  \label{3}
  \scalebox{0.9}{
    \begin{tabular}{|l|rrrrr|}
    \toprule
    \multicolumn{1}{|c|}{\textbf{FTSE100}} & \multicolumn{1}{c}{\textbf{EW}} & \multicolumn{1}{c}{\textbf{MinV}} & \multicolumn{1}{p{5.545em}}{\textbf{MinMAD}} & \multicolumn{1}{c}{\textbf{PT}} & \multicolumn{1}{p{5.545em}|}{\textbf{HF/HE 0.30-0.69}} \\
    \midrule
    \textbf{ExpRet} & \cellcolor[rgb]{ 1,  .922,  .518}0,039\% & \cellcolor[rgb]{ .988,  .749,  .482}0,035\% & \cellcolor[rgb]{ .973,  .412,  .42}0,026\% & \cellcolor[rgb]{ .992,  .922,  .518}0,040\% & \cellcolor[rgb]{ .388,  .745,  .482}0,076\% \\
    \textbf{Vol} & \cellcolor[rgb]{ 1,  .855,  .506}0,013 & \cellcolor[rgb]{ .388,  .745,  .482}0,009 & \cellcolor[rgb]{ .459,  .765,  .486}0,009 & \cellcolor[rgb]{ 1,  .922,  .518}0,011 & \cellcolor[rgb]{ .973,  .412,  .42}0,021 \\
    \textbf{Sharpe} & \cellcolor[rgb]{ .984,  .643,  .463}0,031 & \cellcolor[rgb]{ .388,  .745,  .482}0,038 & \cellcolor[rgb]{ .973,  .412,  .42}0,027 & \cellcolor[rgb]{ 1,  .922,  .518}0,035 & \cellcolor[rgb]{ .988,  .918,  .518}0,035 \\
    \textbf{MaxDD} & \cellcolor[rgb]{ .996,  .863,  .506}-0,487 & \cellcolor[rgb]{ .388,  .745,  .482}-0,353 & \cellcolor[rgb]{ .471,  .769,  .49}-0,366 & \cellcolor[rgb]{ 1,  .922,  .518}-0,459 & \cellcolor[rgb]{ .973,  .412,  .42}-0,712 \\
    \textbf{Sortino} & \cellcolor[rgb]{ .984,  .69,  .471}0,044 & \cellcolor[rgb]{ .388,  .745,  .482}0,053 & \cellcolor[rgb]{ .973,  .412,  .42}0,038 & \cellcolor[rgb]{ 1,  .922,  .518}0,049 & \cellcolor[rgb]{ .675,  .831,  .502}0,051 \\
    \textbf{Rachev} & \cellcolor[rgb]{ .757,  .851,  .506}0,972 & \cellcolor[rgb]{ 1,  .922,  .518}0,938 & \cellcolor[rgb]{ .976,  .541,  .443}0,922 & \cellcolor[rgb]{ .973,  .412,  .42}0,916 & \cellcolor[rgb]{ .388,  .745,  .482}1,023 \\
    \textbf{AveROI} & \cellcolor[rgb]{ 1,  .922,  .518}39\% & \cellcolor[rgb]{ .992,  .831,  .498}37\% & \cellcolor[rgb]{ .973,  .412,  .42}29\% & \cellcolor[rgb]{ .89,  .89,  .514}48\% & \cellcolor[rgb]{ .388,  .745,  .482}108\% \\
    \textbf{NHI} & \multicolumn{1}{c}{-} & \cellcolor[rgb]{ .388,  .745,  .482}0,926 & \cellcolor[rgb]{ .447,  .765,  .486}0,924 & \cellcolor[rgb]{ .996,  .898,  .51}0,895 & \cellcolor[rgb]{ .973,  .412,  .42}0,549 \\
    \textbf{ave \#} & 80    & 21,6  & 20,6  & 29,4  & 4,2 \\
    \bottomrule
    \end{tabular}%
    }%
\end{table}%

On the one hand, from Tables \ref{1}, \ref{2}, and \ref{3}, we can observe that the HF/HE strategy seems to be the most profitable one: it shows the best performances in terms of \textbf{ExpRet}, \textbf{Rachev}, and \textbf{AveROI}, all the datasets considered, and it is superior in terms of \textbf{Sharpe} and \textbf{Sortino} with respect to the other approaches, in the context of DIJA and NASDAQ100 datasets.
On the other hand, as expected, the minimum-risk strategies present the best results in terms of risk, as shown by out-of-sample volatility and \textbf{MaxDD}.
In this sense, the HF/HE strategy seems to be rather aggressive, since it tends to select a handful of stocks.
%
However, as shown in the next section, by appropriately varying $\lambda_+$ and $\lambda_-$, our model appears to be very versatile in that it is able to identify more diversified and less risky portfolios.




\subsection{Flexibility of the HF/HE portfolio selection model}
\label{sec:Flexibility}

By giving a sensitivity analysis, we aim at studying the behavior of the HF/HE approach for different values of $\lambda_+$ and $\lambda_-$, even if they do not necessarily satisfy all paradoxes.  

\subsubsection{A sensitivity analysis: varying HF/HE parameters values\label{sec:calibration}}

The question arises on how the HF/HE approach behaves when the values of parameters $\lambda_+$ and $\lambda_-$ do not necessarily satisfy all the paradoxes presented in \cite{kahnemanandtversky1979} and \cite{birnbaum2008new}.

As mentioned in Section \ref{sec:Preliminary}, the higher the value of $\lambda_+$, the lower the degree of risk-aversion/pessimism, and the higher the value of $\lambda_-$, the higher the degree of risk-seeking/optimism.
More precisely, by decreasing $\lambda_+$, a DM, from being less risk-averse, becomes more risk-averse. Whereas, by decreasing $\lambda_-$, a DM, from more risk-seeking, becomes less risk-seeking.
%
%

\begin{table}[htbp]
  \centering
  \caption{Computational results of the HF/HE approach by varying $\lambda_+$ and $\lambda_-$ on DIJA}
    \label{4}%
  \scalebox{0.8}{
    \begin{tabular}{|c|c|ccc|c|c|c|cc|}
    \toprule
    \multicolumn{10}{|c|}{\textbf{DIJA}} \\
    \midrule
    \multicolumn{2}{|c|}{\multirow{2}[4]{*}{\textbf{ExpRet}}} & \multicolumn{3}{c|}{$\lambda_{+}$} & \multicolumn{2}{c|}{\multirow{2}[4]{*}{\textbf{Vol}}} & \multicolumn{3}{c|}{$\lambda_{+}$} \\
\cmidrule{3-5}\cmidrule{8-10}    \multicolumn{2}{|c|}{} & 0.15  & 0.20  & 0.25  & \multicolumn{2}{c|}{} & \multicolumn{1}{c}{0.15} & 0.20  & 0.25 \\
    \midrule
    \multirow{3}[2]{*}{$\lambda_{-}$} & 0.30  & \cellcolor[rgb]{ .973,  .412,  .42}0,030\% & \cellcolor[rgb]{ .973,  .412,  .42}0,030\% & \cellcolor[rgb]{ .98,  .58,  .451}0,031\% & \multirow{3}[2]{*}{$\lambda_{-}$} & 0.30  & \cellcolor[rgb]{ .388,  .745,  .482}0,0099 & \cellcolor[rgb]{ .588,  .8,  .49}0,0100 & \cellcolor[rgb]{ .792,  .859,  .502}0,0101 \\
          & 0.40  & \cellcolor[rgb]{ .973,  .412,  .42}0,030\% & \cellcolor[rgb]{ 1,  .922,  .518}0,033\% & \cellcolor[rgb]{ .953,  .91,  .518}0,035\% &       & 0.40  & \cellcolor[rgb]{ .792,  .859,  .502}0,0101 & \cellcolor[rgb]{ 1,  .922,  .518}0,0102 & \cellcolor[rgb]{ 1,  .894,  .514}0,0104 \\
          & 0.66  & \cellcolor[rgb]{ .929,  .902,  .514}0,036\% & \cellcolor[rgb]{ .855,  .882,  .51}0,039\% & \cellcolor[rgb]{ .388,  .745,  .482}0,058\% &       & 0.66  & \cellcolor[rgb]{ .992,  .773,  .49}0,0113 & \cellcolor[rgb]{ .988,  .635,  .463}0,0123 & \cellcolor[rgb]{ .973,  .412,  .42}0,0139 \\
    \midrule
    \multicolumn{2}{|c|}{\multirow{2}[4]{*}{\textbf{Sharpe}}} & \multicolumn{3}{c|}{$\lambda_{+}$} & \multicolumn{2}{c|}{\multirow{2}[4]{*}{\textbf{ave \#}}} & \multicolumn{3}{c|}{$\lambda\_{+}$} \\
\cmidrule{3-5}\cmidrule{8-10}    \multicolumn{2}{|c|}{} & 0.15  & 0.20  & 0.25  & \multicolumn{2}{c|}{} & \multicolumn{1}{c}{0.15} & 0.20  & 0.25 \\
    \midrule
    \multirow{3}[2]{*}{$\lambda_{-}$} & 0.30  & \cellcolor[rgb]{ .988,  .718,  .478}0,0307 & \cellcolor[rgb]{ .973,  .412,  .42}0,0298 & \cellcolor[rgb]{ .976,  .51,  .435}0,0301 & \multirow{3}[2]{*}{$\lambda_{-}$} & 0.30  & \cellcolor[rgb]{ .514,  .78,  .49}10,8 & \cellcolor[rgb]{ .635,  .82,  .498}10,6 & \cellcolor[rgb]{ 1,  .922,  .518}10,0 \\
          & 0.40  & \cellcolor[rgb]{ .976,  .51,  .435}0,0301 & \cellcolor[rgb]{ .965,  .914,  .518}0,0319 & \cellcolor[rgb]{ .851,  .878,  .51}0,0338 &       & 0.40  & \cellcolor[rgb]{ .388,  .745,  .482}11,0 & \cellcolor[rgb]{ .878,  .89,  .514}10,2 & \cellcolor[rgb]{ .996,  .867,  .506}9,6 \\
          & 0.66  & \cellcolor[rgb]{ .949,  .906,  .518}0,0322 & \cellcolor[rgb]{ 1,  .922,  .518}0,0313 & \cellcolor[rgb]{ .388,  .745,  .482}0,0414 &       & 0.66  & \cellcolor[rgb]{ .996,  .894,  .51}9,8 & \cellcolor[rgb]{ .984,  .682,  .471}8,2 & \cellcolor[rgb]{ .973,  .412,  .42}6,1 \\
    \bottomrule
    \end{tabular}}%
\end{table}%

\begin{table}[htbp]
  \centering
  \caption{Computational results of the HF/HE approach by varying $\lambda_+$ and $\lambda_-$ on NASDAQ100}
  \label{5}%
  \scalebox{0.8}{
    \begin{tabular}{|c|c|ccc|c|c|c|cc|}
    \toprule
    \multicolumn{10}{|c|}{\textbf{NASDAQ100}} \\
    \midrule
    \multicolumn{2}{|c|}{\multirow{2}[4]{*}{\textbf{ExpRet}}} & \multicolumn{3}{c|}{$\lambda_{+}$} & \multicolumn{2}{c|}{\multirow{2}[4]{*}{\textbf{Vol}}} & \multicolumn{3}{c|}{$\lambda_{+}$} \\
\cmidrule{3-5}\cmidrule{8-10}    \multicolumn{2}{|c|}{} & 0.15  & 0.20  & 0.25  & \multicolumn{2}{c|}{} & \multicolumn{1}{c}{0.15} & 0.20  & 0.25 \\
    \midrule
    \multirow{3}[2]{*}{$\lambda_{-}$} & 0.30  & \cellcolor[rgb]{ .973,  .412,  .42}0,041\% & \cellcolor[rgb]{ .973,  .412,  .42}0,041\% & \cellcolor[rgb]{ .984,  .667,  .467}0,043\% & \multirow{3}[2]{*}{$\lambda_{-}$} & 0.30  & \cellcolor[rgb]{ .388,  .745,  .482}0,0105 & \cellcolor[rgb]{ .631,  .812,  .494}0,0107 & \cellcolor[rgb]{ 1,  .922,  .518}0,0110 \\
          & 0.40  & \cellcolor[rgb]{ .976,  .537,  .443}0,042\% & \cellcolor[rgb]{ 1,  .922,  .518}0,045\% & \cellcolor[rgb]{ .98,  .918,  .518}0,047\% &       & 0.40  & \cellcolor[rgb]{ .631,  .812,  .494}0,0107 & \cellcolor[rgb]{ 1,  .922,  .518}0,0110 & \cellcolor[rgb]{ 1,  .894,  .514}0,0114 \\
          & 0.66  & \cellcolor[rgb]{ .855,  .882,  .51}0,060\% & \cellcolor[rgb]{ .655,  .824,  .498}0,080\% & \cellcolor[rgb]{ .388,  .745,  .482}0,107\% &       & 0.66  & \cellcolor[rgb]{ .996,  .792,  .494}0,0128 & \cellcolor[rgb]{ .984,  .627,  .463}0,0150 & \cellcolor[rgb]{ .973,  .412,  .42}0,0179 \\
    \midrule
    \multicolumn{2}{|c|}{\multirow{2}[4]{*}{\textbf{Sharpe}}} & \multicolumn{3}{c|}{$\lambda_{+}$} & \multicolumn{2}{c|}{\multirow{2}[4]{*}{\textbf{ave \#}}} & \multicolumn{3}{c|}{$\lambda_{+}$} \\
\cmidrule{3-5}\cmidrule{8-10}    \multicolumn{2}{|c|}{} & 0.15  & 0.20  & 0.25  & \multicolumn{2}{c|}{} & \multicolumn{1}{c}{0.15} & 0.20  & 0.25 \\
    \midrule
    \multirow{3}[2]{*}{$\lambda_{-}$} & 0.30  & \cellcolor[rgb]{ .98,  .58,  .451}0,0393 & \cellcolor[rgb]{ .976,  .557,  .447}0,0392 & \cellcolor[rgb]{ .98,  .627,  .459}0,0395 & \multirow{3}[2]{*}{$\lambda_{-}$} & 0.30  & \cellcolor[rgb]{ .467,  .769,  .49}14,6 & \cellcolor[rgb]{ .506,  .78,  .49}14,5 & \cellcolor[rgb]{ 1,  .922,  .518}13,2 \\
          & 0.40  & \cellcolor[rgb]{ .973,  .412,  .42}0,0386 & \cellcolor[rgb]{ 1,  .922,  .518}0,0407 & \cellcolor[rgb]{ .98,  .918,  .518}0,0414 &       & 0.40  & \cellcolor[rgb]{ .388,  .745,  .482}14,8 & \cellcolor[rgb]{ .812,  .867,  .51}13,7 & \cellcolor[rgb]{ .996,  .859,  .502}12,5 \\
          & 0.66  & \cellcolor[rgb]{ .808,  .867,  .51}0,0467 & \cellcolor[rgb]{ .596,  .808,  .498}0,0534 & \cellcolor[rgb]{ .388,  .745,  .482}0,0598 &       & 0.66  & \cellcolor[rgb]{ 1,  .922,  .518}13,2 & \cellcolor[rgb]{ .984,  .698,  .475}10,7 & \cellcolor[rgb]{ .973,  .412,  .42}7,4 \\
    \bottomrule
    \end{tabular}}%
\end{table}%

\begin{table}[htbp]
  \centering
    \caption{Computational results of the HF/HE approach by varying $\lambda_+$ and $\lambda_-$ on FTSE100}
    \label{6}%
    \scalebox{0.8}{
    \begin{tabular}{|c|c|ccc|c|c|c|cc|}
    \toprule
    \multicolumn{10}{|c|}{\textbf{FTSE 100}} \\
    \midrule
    \multicolumn{2}{|c|}{\multirow{2}[4]{*}{\textbf{ExpRet}}} & \multicolumn{3}{c|}{$\lambda_{+}$} & \multicolumn{2}{c|}{\multirow{2}[4]{*}{\textbf{Vol}}} & \multicolumn{3}{c|}{$\lambda_{+}$} \\
\cmidrule{3-5}\cmidrule{8-10}    \multicolumn{2}{|c|}{} & 0.15  & 0.20  & 0.25  & \multicolumn{2}{c|}{} & \multicolumn{1}{c}{0.15} & 0.20  & 0.25 \\
    \midrule
    \multirow{3}[2]{*}{$\lambda_{-}$} & 0.30  & \cellcolor[rgb]{ .973,  .412,  .42}0,036\% & \cellcolor[rgb]{ .984,  .663,  .467}0,037\% & \cellcolor[rgb]{ 1,  .922,  .518}0,038\% & \multirow{3}[2]{*}{$\lambda_{-}$} & 0.30  & \cellcolor[rgb]{ .388,  .745,  .482}0,0100 & \cellcolor[rgb]{ .537,  .788,  .49}0,0101 & \cellcolor[rgb]{ 1,  .922,  .518}0,0104 \\
          & 0.40  & \cellcolor[rgb]{ .973,  .412,  .42}0,036\% & \cellcolor[rgb]{ .984,  .663,  .467}0,037\% & \cellcolor[rgb]{ 1,  .922,  .518}0,038\% &       & 0.40  & \cellcolor[rgb]{ .694,  .831,  .498}0,0102 & \cellcolor[rgb]{ 1,  .922,  .518}0,0104 & \cellcolor[rgb]{ 1,  .886,  .514}0,0107 \\
          & 0.66  & \cellcolor[rgb]{ .937,  .906,  .518}0,040\% & \cellcolor[rgb]{ .584,  .804,  .494}0,051\% & \cellcolor[rgb]{ .388,  .745,  .482}0,057\% &       & 0.66  & \cellcolor[rgb]{ .996,  .788,  .494}0,0115 & \cellcolor[rgb]{ .988,  .643,  .467}0,0127 & \cellcolor[rgb]{ .973,  .412,  .42}0,0146 \\
    \midrule
    \multicolumn{2}{|c|}{\multirow{2}[4]{*}{\textbf{Sharpe}}} & \multicolumn{3}{c|}{$\lambda_{+}$} & \multicolumn{2}{c|}{\multirow{2}[4]{*}{\textbf{ave \#}}} & \multicolumn{3}{c|}{$\lambda_{+}$} \\
\cmidrule{3-5}\cmidrule{8-10}    \multicolumn{2}{|c|}{} & 0.15  & 0.20  & 0.25  & \multicolumn{2}{c|}{} & \multicolumn{1}{c}{0.15} & 0.20  & 0.25 \\
    \midrule
    \multirow{3}[2]{*}{$\lambda_{-}$} & 0.30  & \cellcolor[rgb]{ .961,  .91,  .518}0,0363 & \cellcolor[rgb]{ 1,  .922,  .518}0,0360 & \cellcolor[rgb]{ .859,  .882,  .51}0,0370 & \multirow{3}[2]{*}{$\lambda_{-}$} & 0.30  & \cellcolor[rgb]{ .388,  .745,  .482}18,3 & \cellcolor[rgb]{ .455,  .765,  .486}18,1 & \cellcolor[rgb]{ 1,  .922,  .518}16,4 \\
          & 0.40  & \cellcolor[rgb]{ .992,  .82,  .498}0,0358 & \cellcolor[rgb]{ .973,  .459,  .427}0,0351 & \cellcolor[rgb]{ .988,  .769,  .486}0,0357 &       & 0.40  & \cellcolor[rgb]{ .424,  .757,  .486}18,2 & \cellcolor[rgb]{ .937,  .906,  .518}16,6 & \cellcolor[rgb]{ .992,  .835,  .498}15,2 \\
          & 0.66  & \cellcolor[rgb]{ .973,  .412,  .42}0,0350 & \cellcolor[rgb]{ .388,  .745,  .482}0,0403 & \cellcolor[rgb]{ .588,  .804,  .494}0,0389 &       & 0.66  & \cellcolor[rgb]{ .996,  .855,  .502}15,5 & \cellcolor[rgb]{ .984,  .671,  .467}12,9 & \cellcolor[rgb]{ .973,  .412,  .42}9,2 \\
    \bottomrule
    \end{tabular}}%
\end{table}%

\noindent In Tables \ref{4}, \ref{5}, and \ref{6}, we report the results in terms of \textbf{ExpRet}, \textbf{Vol}, \textbf{Sharpe}, and \textbf{ave \#}, that we obtain for several HF/HE portfolios, by setting different values of $\lambda_+$ and $\lambda_-$.  

Interestingly, on the one hand, by simultaneously decreasing $\lambda_+$ and $\lambda_-$, the corresponding HF/HE portfolios tend to show a reduction in terms of volatility and an increase in terms of the number of selected assets: values of \textbf{Vol} and \textbf{ave \#} turn out to be comparable with those obtained through the minimum risk strategies.
On the other hand, by simultaneously increasing $\lambda_+$ and $\lambda_-$, the HF/HE portfolios tend to exhibit an increase in terms of \textbf{ExpRet} and \textbf{Sharpe}.

\subsubsection{Comparing HF/HE portfolios\label{sec:HF/HEportfolios}}

It is appropriate to compare the out-of-sample performances of the HF/HE portfolio whose parameters satisfy all the experiments of \cite{kahnemanandtversky1979} and \cite{birnbaum2008new} (i.e., $\lambda_+=0.30$ and $\lambda_-=0.69$) with those of several HF/HE portfolios that are obtained corresponding to different values of $\lambda_+$ and $\lambda_-$. Hence, we consider the following selection strategies:
\begin{itemize}
    \item HF/HE 0.25-0.66: Model \eqref{hfhe_fmincon} with $\lambda_+=0.25$ and $\lambda_-=0.66$;
    \item HF/HE 0.25-0.40: Model \eqref{hfhe_fmincon} with $\lambda_+=0.25$ and $\lambda_-=0.40$;
    \item HF/HE 0.20-0.66: Model \eqref{hfhe_fmincon} with $\lambda_+=0.20$ and $\lambda_-=0.66$;
    \item HF/HE 0.20-0.40: Model \eqref{hfhe_fmincon} with $\lambda_+=0.20$ and $\lambda_-=0.40$; 
    \item HF/HE 0.15-0.30: Model \eqref{hfhe_fmincon} with $\lambda_+=0.15$ and $\lambda_-=0.30$.
\end{itemize}
In Tables \ref{7}, \ref{8}, and \ref{9}, we indicate the out-of-sample performance results for the HF/HE portfolios listed above. 
Note that the different values adopted for $\lambda_{+}$ and $\lambda_{-}$ are marked in pink and blue, respectively.

On the one hand, we can observe again that the HF/HE portfolio with $\lambda_+=0.30$ and $\lambda_-=0.69$ appears to yield the most profitable strategy, achieving the best performances in terms of \textbf{ExpRet}, \textbf{Rachev}, and \textbf{AveROI}, all the datasets considered.
On the other hand, choosing smaller values for parameters $\lambda_{+}$ and $\lambda_{-}$, we obtain HF/HE portfolios that are very competitive in terms of risk and diversification with respect to PT and minimum risk strategies.

\begin{table}[htbp]
  \centering
  \caption{Computational results on DIJA}
    \label{7}
      \scalebox{0.8}{
    \begin{tabular}{|l|rrrrrr|}
    \toprule
    \multicolumn{1}{|c|}{\textbf{DIJA}} & \multicolumn{1}{p{4.50em}}{\textbf{HF/HE \textcolor[rgb]{1.00,0.50,1.00}{0.30}-\textcolor[rgb]{0.00,0.50,1.00}{0.69}}} & \multicolumn{1}{p{4.5em}}{\textbf{HF/HE \textcolor[rgb]{1.00,0.50,1.00}{0.25}-\textcolor[rgb]{0.00,0.50,1.00}{0.66}}} & \multicolumn{1}{p{4.50em}}{\textbf{HF/HE \textcolor[rgb]{1.00,0.50,1.00}{0.25}-\textcolor[rgb]{0.00,0.50,1.00}{0.40}}} & \multicolumn{1}{p{4.5em}}{\textbf{HF/HE \textcolor[rgb]{1.00,0.50,1.00}{0.20}-\textcolor[rgb]{0.00,0.50,1.00}{0.66}}} & \multicolumn{1}{p{4.5em}}{\textbf{HF/HE \textcolor[rgb]{1.00,0.50,1.00}{0.20}-\textcolor[rgb]{0.00,0.50,1.00}{0.40}}} & \multicolumn{1}{p{4.5em}|}{\textbf{HF/HE \textcolor[rgb]{1.00,0.50,1.00}{0.15}-\textcolor[rgb]{0.00,0.50,1.00}{0.30}}} \\
    \midrule
    \textbf{ExpRet} & \cellcolor[rgb]{ .388,  .745,  .482}0,073\% & \cellcolor[rgb]{ .647,  .824,  .498}0,058\% & \cellcolor[rgb]{ .992,  .788,  .49}0,035\% & \cellcolor[rgb]{ .973,  .914,  .518}0,039\% & \cellcolor[rgb]{ .98,  .584,  .451}0,033\% & \cellcolor[rgb]{ .973,  .412,  .42}0,030\% \\
    \textbf{Vol} & \cellcolor[rgb]{ .973,  .412,  .42}0,018 & \cellcolor[rgb]{ .992,  .729,  .482}0,014 & \cellcolor[rgb]{ .612,  .808,  .494}0,010 & \cellcolor[rgb]{ .996,  .851,  .506}0,012 & \cellcolor[rgb]{ .518,  .78,  .486}0,010 & \cellcolor[rgb]{ .388,  .745,  .482}0,010 \\
    \textbf{Sharpe} & \cellcolor[rgb]{ .463,  .769,  .49}0,040 & \cellcolor[rgb]{ .388,  .745,  .482}0,041 & \cellcolor[rgb]{ .937,  .906,  .518}0,034 & \cellcolor[rgb]{ .98,  .561,  .447}0,031 & \cellcolor[rgb]{ .988,  .706,  .475}0,032 & \cellcolor[rgb]{ .973,  .412,  .42}0,031 \\
    \textbf{MaxDD} & \cellcolor[rgb]{ .973,  .412,  .42}-0,580 & \cellcolor[rgb]{ .996,  .855,  .502}-0,404 & \cellcolor[rgb]{ .506,  .78,  .49}-0,351 & \cellcolor[rgb]{ .992,  .82,  .498}-0,417 & \cellcolor[rgb]{ .388,  .745,  .482}-0,345 & \cellcolor[rgb]{ .525,  .788,  .494}-0,353 \\
    \textbf{Sortino} & \cellcolor[rgb]{ .388,  .745,  .482}0,060 & \cellcolor[rgb]{ .416,  .753,  .486}0,059 & \cellcolor[rgb]{ .945,  .906,  .518}0,048 & \cellcolor[rgb]{ .976,  .518,  .439}0,044 & \cellcolor[rgb]{ .988,  .71,  .475}0,045 & \cellcolor[rgb]{ .973,  .412,  .42}0,043 \\
    \textbf{Rachev} & \cellcolor[rgb]{ .388,  .745,  .482}1,030 & \cellcolor[rgb]{ .718,  .843,  .502}0,979 & \cellcolor[rgb]{ .992,  .792,  .49}0,931 & \cellcolor[rgb]{ .98,  .918,  .518}0,937 & \cellcolor[rgb]{ .988,  .757,  .486}0,930 & \cellcolor[rgb]{ .973,  .412,  .42}0,922 \\
    \textbf{AveROI} & \cellcolor[rgb]{ .388,  .745,  .482}65\% & \cellcolor[rgb]{ .678,  .831,  .502}51\% & \cellcolor[rgb]{ .996,  .882,  .51}35\% & \cellcolor[rgb]{ .996,  .922,  .518}36\% & \cellcolor[rgb]{ .976,  .486,  .431}31\% & \cellcolor[rgb]{ .973,  .412,  .42}30\% \\
    \textbf{NHI} & \cellcolor[rgb]{ .973,  .412,  .42} 0,427 & \cellcolor[rgb]{ .992,  .788,  .49}0,716 & \cellcolor[rgb]{ .761,  .855,  .506}0,827 & \cellcolor[rgb]{ .996,  .906,  .514}0,806 & \cellcolor[rgb]{ .459,  .769,  .49}0,840 & \cellcolor[rgb]{ .388,  .745,  .482}0,843 \\
    \textbf{ave \#} & 2,9   & 6,1   & 9,6   & 8,2   & 10,2  & 10,8 \\
    \bottomrule
    \end{tabular}%
  }
\end{table}%

\begin{table}[htbp]
  \centering
   \caption{Computational results on NASDAQ100}
  \label{8}
      \scalebox{0.8}{
    \begin{tabular}{|l|rrrrrr|}
    \toprule
    \multicolumn{1}{|c|}{\textbf{NASDAQ100}} & \multicolumn{1}{p{4.50em}}{\textbf{HF/HE \textcolor[rgb]{1.00,0.50,1.00}{0.30}-\textcolor[rgb]{0.00,0.50,1.00}{0.69}}} & \multicolumn{1}{p{4.5em}}{\textbf{HF/HE \textcolor[rgb]{1.00,0.50,1.00}{0.25}-\textcolor[rgb]{0.00,0.50,1.00}{0.66}}} & \multicolumn{1}{p{4.50em}}{\textbf{HF/HE \textcolor[rgb]{1.00,0.50,1.00}{0.25}-\textcolor[rgb]{0.00,0.50,1.00}{0.40}}} & \multicolumn{1}{p{4.5em}}{\textbf{HF/HE \textcolor[rgb]{1.00,0.50,1.00}{0.20}-\textcolor[rgb]{0.00,0.50,1.00}{0.66}}} & \multicolumn{1}{p{4.5em}}{\textbf{HF/HE \textcolor[rgb]{1.00,0.50,1.00}{0.20}-\textcolor[rgb]{0.00,0.50,1.00}{0.40}}} & \multicolumn{1}{p{4.5em}|}{\textbf{HF/HE \textcolor[rgb]{1.00,0.50,1.00}{0.15}-\textcolor[rgb]{0.00,0.50,1.00}{0.30}}} \\
    \midrule
    \textbf{ExpRet} & \cellcolor[rgb]{ .388,  .745,  .482}0,141\% & \cellcolor[rgb]{ .659,  .824,  .498}0,107\% & \cellcolor[rgb]{ .976,  .545,  .443}0,047\% & \cellcolor[rgb]{ .875,  .886,  .514}0,080\% & \cellcolor[rgb]{ .976,  .49,  .431}0,045\% & \cellcolor[rgb]{ .973,  .412,  .42}0,041\% \\
    \textbf{Vol} & \cellcolor[rgb]{ .973,  .412,  .42}0,024 & \cellcolor[rgb]{ .992,  .71,  .478}0,018 & \cellcolor[rgb]{ .592,  .804,  .494}0,011 & \cellcolor[rgb]{ .996,  .843,  .506}0,015 & \cellcolor[rgb]{ .498,  .776,  .486}0,011 & \cellcolor[rgb]{ .388,  .745,  .482}0,011 \\
    \textbf{Sharpe} & \cellcolor[rgb]{ .49,  .776,  .49}0,058 & \cellcolor[rgb]{ .388,  .745,  .482}0,060 & \cellcolor[rgb]{ .976,  .541,  .443}0,041 & \cellcolor[rgb]{ .706,  .839,  .502}0,053 & \cellcolor[rgb]{ .976,  .498,  .435}0,041 & \cellcolor[rgb]{ .973,  .412,  .42}0,039 \\
    \textbf{MaxDD} & \cellcolor[rgb]{ .973,  .412,  .42}-0,582 & \cellcolor[rgb]{ .992,  .839,  .502}-0,442 & \cellcolor[rgb]{ .925,  .902,  .514}-0,397 & \cellcolor[rgb]{ .996,  .906,  .514}-0,411 & \cellcolor[rgb]{ .592,  .804,  .494}-0,365 & \cellcolor[rgb]{ .388,  .745,  .482}-0,346 \\
    \textbf{Sortino} & \cellcolor[rgb]{ .459,  .769,  .49}0,086 & \cellcolor[rgb]{ .388,  .745,  .482}0,089 & \cellcolor[rgb]{ .976,  .525,  .439}0,058 & \cellcolor[rgb]{ .729,  .843,  .502}0,077 & \cellcolor[rgb]{ .976,  .486,  .431}0,057 & \cellcolor[rgb]{ .973,  .412,  .42}0,055 \\
    \textbf{Rachev} & \cellcolor[rgb]{ .388,  .745,  .482}1,054 & \cellcolor[rgb]{ .537,  .788,  .494}1,028 & \cellcolor[rgb]{ .973,  .412,  .42}0,914 & \cellcolor[rgb]{ .851,  .878,  .51}0,974 & \cellcolor[rgb]{ .973,  .455,  .427}0,917 & \cellcolor[rgb]{ .976,  .514,  .439}0,921 \\
    \textbf{AveROI} & \cellcolor[rgb]{ .388,  .745,  .482}233\% & \cellcolor[rgb]{ .706,  .839,  .502}153\% & \cellcolor[rgb]{ .98,  .584,  .451}57\% & \cellcolor[rgb]{ .922,  .902,  .514}98\% & \cellcolor[rgb]{ .976,  .51,  .435}52\% & \cellcolor[rgb]{ .973,  .412,  .42}46\% \\
    \textbf{NHI} & \cellcolor[rgb]{ .973,  .412,  .42} 0,448 & \cellcolor[rgb]{ .992,  .776,  .486}0,726 & \cellcolor[rgb]{ .643,  .82,  .498}0,851 & \cellcolor[rgb]{ .996,  .898,  .514}0,820 & \cellcolor[rgb]{ .482,  .773,  .49}0,858 & \cellcolor[rgb]{ .388,  .745,  .482}0,862 \\
    \textbf{ave \#} & 3,6   & 7,4   & 12,5  & 10,7  & 13,7  & 14,6 \\
    \bottomrule
    \end{tabular}%
  }
\end{table}%

\begin{table}[htbp]
  \centering
  \caption{Computational results on FTSE100}
  \label{9}
      \scalebox{0.8}{
    \begin{tabular}{|l|rrrrrr|}
    \toprule
    \multicolumn{1}{|c|}{\textbf{FTSE 100}} & \multicolumn{1}{p{4.50em}}{\textbf{HF/HE \textcolor[rgb]{1.00,0.50,1.00}{0.30}-\textcolor[rgb]{0.00,0.50,1.00}{0.69}}} & \multicolumn{1}{p{4.5em}}{\textbf{HF/HE \textcolor[rgb]{1.00,0.50,1.00}{0.25}-\textcolor[rgb]{0.00,0.50,1.00}{0.66}}} & \multicolumn{1}{p{4.50em}}{\textbf{HF/HE \textcolor[rgb]{1.00,0.50,1.00}{0.25}-\textcolor[rgb]{0.00,0.50,1.00}{0.40}}} & \multicolumn{1}{p{4.5em}}{\textbf{HF/HE \textcolor[rgb]{1.00,0.50,1.00}{0.20}-\textcolor[rgb]{0.00,0.50,1.00}{0.66}}} & \multicolumn{1}{p{4.5em}}{\textbf{HF/HE \textcolor[rgb]{1.00,0.50,1.00}{0.20}-\textcolor[rgb]{0.00,0.50,1.00}{0.40}}} & \multicolumn{1}{p{4.5em}|}{\textbf{HF/HE \textcolor[rgb]{1.00,0.50,1.00}{0.15}-\textcolor[rgb]{0.00,0.50,1.00}{0.30}}} \\
    \midrule
    \textbf{ExpRet} & \cellcolor[rgb]{ .388,  .745,  .482}0,076\% & \cellcolor[rgb]{ .765,  .855,  .506}0,057\% & \cellcolor[rgb]{ .976,  .533,  .443}0,038\% & \cellcolor[rgb]{ .878,  .886,  .514}0,051\% & \cellcolor[rgb]{ .973,  .427,  .42}0,037\% & \cellcolor[rgb]{ .973,  .412,  .42}0,036\% \\
    \textbf{Vol} & \cellcolor[rgb]{ .973,  .412,  .42}0,021 & \cellcolor[rgb]{ .992,  .773,  .49}0,015 & \cellcolor[rgb]{ .647,  .82,  .494}0,011 & \cellcolor[rgb]{ 1,  .871,  .51}0,013 & \cellcolor[rgb]{ .541,  .788,  .49}0,010 & \cellcolor[rgb]{ .388,  .745,  .482}0,010 \\
    \textbf{Sharpe} & \cellcolor[rgb]{ .98,  .62,  .459}0,035 & \cellcolor[rgb]{ .584,  .804,  .494}0,039 & \cellcolor[rgb]{ .988,  .773,  .486}0,036 & \cellcolor[rgb]{ .388,  .745,  .482}0,040 & \cellcolor[rgb]{ .973,  .412,  .42}0,035 & \cellcolor[rgb]{ .965,  .914,  .518}0,036 \\
    \textbf{MaxDD} & \cellcolor[rgb]{ .973,  .412,  .42}-0,712 & \cellcolor[rgb]{ .992,  .773,  .486}-0,538 & \cellcolor[rgb]{ .486,  .776,  .49}-0,431 & \cellcolor[rgb]{ .992,  .839,  .502}-0,507 & \cellcolor[rgb]{ .416,  .753,  .486}-0,426 & \cellcolor[rgb]{ .388,  .745,  .482}-0,424 \\
    \textbf{Sortino} & \cellcolor[rgb]{ .918,  .898,  .514}0,051 & \cellcolor[rgb]{ .49,  .776,  .49}0,055 & \cellcolor[rgb]{ .988,  .745,  .482}0,049 & \cellcolor[rgb]{ .388,  .745,  .482}0,056 & \cellcolor[rgb]{ .973,  .412,  .42}0,048 & \cellcolor[rgb]{ 1,  .922,  .518}0,050 \\
    \textbf{Rachev} & \cellcolor[rgb]{ .388,  .745,  .482}1,023 & \cellcolor[rgb]{ .816,  .871,  .51}0,950 & \cellcolor[rgb]{ .98,  .557,  .447}0,903 & \cellcolor[rgb]{ .91,  .898,  .514}0,934 & \cellcolor[rgb]{ .973,  .459,  .427}0,899 & \cellcolor[rgb]{ .973,  .412,  .42}0,896 \\
    \textbf{AveROI} & \cellcolor[rgb]{ .388,  .745,  .482}108\% & \cellcolor[rgb]{ .78,  .859,  .506}76\% & \cellcolor[rgb]{ .98,  .58,  .451}49\% & \cellcolor[rgb]{ .902,  .894,  .514}66\% & \cellcolor[rgb]{ .973,  .463,  .427}47\% & \cellcolor[rgb]{ .973,  .412,  .42}45\% \\
    \textbf{NHI} & \cellcolor[rgb]{ .973,  .412,  .42} 0,549 & \cellcolor[rgb]{ .992,  .776,  .49}0,785 & \cellcolor[rgb]{ .718,  .843,  .502}0,892 & \cellcolor[rgb]{ .996,  .894,  .51}0,859 & \cellcolor[rgb]{ .561,  .796,  .494}0,901 & \cellcolor[rgb]{ .388,  .745,  .482}0,911 \\
    \textbf{ave \#} & 4,2   & 9,2   & 15,2  & 12,9  & 16,6  & 18,3 \\
    \bottomrule
    \end{tabular}%
  }
\end{table}%

\section{Conclusions}\label{sec:conclusions}

Taking a cue from \cite{cencietal2015} and \cite{corradini2022half}, we have extended the so-called HF/HE behavioral approach to encompass the general case of mixed positive and negative lotteries. Also, we have been able to present a novel and detailed discussion about the financial meaning of the model's relevant parameters.
Such analysis, in turn, prompted us to tailor this strategy to portfolio selection purposes: as a result, we devised a nonconvex optimization model together with its MILP reformulation.
Concerning testing and validation, we have presented an extensive empirical analysis based on several real-world datasets, which confirms, by means of numerical evidences, the out-of-sample competitiveness of the method.

Future research will be directed toward (a) a better understanding of the role played by the probabilistic bias parameter $q$ in portfolio selection, (b) a thorough investigation of different algorithmic strategies to address the problem in practice, and (c) a ``behavioral evaluation'' of financial derivative contracts.

{\footnotesize
\bibliographystyle{apa}
\bibliography{Bib_CCLR_HFHE_202031115}
}

\end{document}